\documentclass[pra,aps,superscriptaddress,amsmath,amssymb,twocolumn,floatfix,longbibliography]{revtex4-1}
\usepackage[pdftex]{graphicx}
\usepackage{bm}
\usepackage{bbold}
\usepackage{dsfont}
\usepackage[colorlinks,linkcolor=blue,citecolor=blue,urlcolor=blue]{hyperref}
\usepackage[usenames,dvipsnames]{color}

\newcommand{\bg}{\begin{gather} }
\newcommand{\eg}{\end{gather}}

\newcommand{\be}{\begin{equation} }
\newcommand{\ee}{\end{equation}}

\newcommand{\corr}[1]{\langle #1\rangle}

\newcommand{\str}{\mathop{\rm str}}
\newcommand{\sdet}{\mathop{\rm sdet}}

\newcommand{\tr}{\mathop{\rm tr}}
\newcommand{\diag}{\mathop{\rm diag}}

\newcommand{\sign}{\mathop{\rm sign}}
\newcommand{\var}{\mathop{\rm var}}
\newcommand{\res}{\mathop{\rm res}}

\newcommand{\thone}{\theta_\text{B1}}
\newcommand{\thtwo}{\theta_\text{B2}}
\newcommand{\thF}{\theta_\text{F}}

\newcommand{\aone}{a_\text{B1}}
\newcommand{\atwo}{a_\text{B2}}
\newcommand{\aF}{a_\text{F}}

\newcommand{\gavg}{\langle g \rangle}
\newcommand{\xiloc}{\xi_{\text{loc}}}

\def\black{\color{black}}

\begin{document}

\title{Mesoscopic conductance fluctuations and noise in disordered Majorana wires}

\author{Daniil S. Antonenko}
\affiliation{Skolkovo Institute of Science and Technology, Moscow 121205, Russia}
\affiliation{L. D. Landau Institute for Theoretical Physics, Chernogolovka 142432, Russia}
\affiliation{Moscow Institute of Physics and Technology, Moscow 141700, Russia}

\author{Eslam Khalaf}
\affiliation{Department of Physics, Harvard University, Cambridge, MA 02138, USA}

\author{Pavel M. Ostrovsky}
\affiliation{Max-Planck-Institut f{\"u}r Festk{\"o}rperforschung, 70569 Stuttgart, Germany}
\affiliation{L. D. Landau Institute for Theoretical Physics, Chernogolovka 142432, Russia}

\author{Mikhail A. Skvortsov}
\affiliation{Skolkovo Institute of Science and Technology, Moscow 121205, Russia}
\affiliation{L. D. Landau Institute for Theoretical Physics, Chernogolovka 142432, Russia}

\date{November 30, 2020}

\begin{abstract}
Superconducting wires with broken time-reversal and spin-rotational symmetries can exhibit two distinct topological gapped phases and host bound Majorana states at the phase boundaries. When the wire is tuned to the transition between these two phases and the gap is closed, Majorana states become delocalized leading to a peculiar critical state of the system. We study transport properties of this critical state as a function of the length $L$
of a disordered multichannel wire.
Applying a non-linear
supersymmetric sigma model of symmetry class D with two replicas, we identify the average conductance,
its variance  and the third cumulant in the whole range of $L$ from the Ohmic limit of short wires to the regime of a broad conductance distribution when $L$ exceeds the correlation length of the system. In addition, we calculate the average thermal shot noise power and variance of the topological index for arbitrary $L$. The general approach developed in the paper can also be applied to study combined effects of disorder and topology in wires of other symmetries.
\end{abstract}

\maketitle

\section{Introduction}

Topological insulators hosting gapless excitations at their boundary have been the subject of intense studies during the last two decades \cite{HasanKane}. One of the most fascinating features of topological materials is the possibility of observing Majorana fermions in a solid state setup. Manipulation of topologically protected Majorana bound states is believed to be a promising platform for quantum computation and information processing, since it may overcome the decoherence issue in  conventional qubits \cite{Beenakker-Majorana}. Several realizations of Majorana fermions have been proposed, including vortex bound states in $p$-wave superconductors \cite{Ivanov2001}, Kitaev chain \cite{Kitaev_chain}, and a semiconducting wire proximized with an $s$-wave superconductor \cite{Lutchyn,Oreg}. The latter approach turned out to be the most suitable for experimental implementation, and a number of publications reporting observation of Majorana states in proximized quantum wires have appeared in recent years \cite{Mourik2012, Das2012, Kouwenhoven, Xu_experiment_2015, Lv_experiment_2017, Chen_experiment_2017, Lutchyn_review_2018}.

Majorana fermion formation requires a superconductor with broken time-reversal and spin-rotation symmetries, a system of symmetry class D in the classification of Ref.~\onlinecite{AZ}. In one spatial dimension (quantum wires), this class is characterized by $\mathds{Z}_2$ topological quantum number \cite{10foldway,Kitaev2009}, indicating that there exist two topologically distinct phases.
Both trivial ($q=1$) and topological ($q=-1$) phases have a spectral gap, which by tuning a control parameter $\mu$ is closed and then reopens with a different sign, giving rise to a Majorana mode localized at the phase boundary. In a clean, translationally invariant system, the topological invariant $q$ can be expressed in terms of the Pfaffians of the Hamiltonian in the center and at the corner of the Brillouin zone \cite{Kitaev_chain}.

Disorder, inevitably present in experimental realizations, affects the above picture in several ways. First, it may shift the position of the  border between the topological phases \cite{BagretsKamenev}. Second, it breaks translational invariance, compromising classification of topological phases in the momentum representation. Nevertheless, \emph{for a given disorder realization} it is still possible to distinguish topological phases by analyzing the real-space transport properties. The topological invariant for class-D quantum wires can be expressed in terms of the matrix $r$ of quasiparticle reflection amplitudes as $q=\sign\det r$ \cite{Akhmerov_classD}. Hence, right at the transition between the topological phases, the wire has a fully open channel with unit transmission.

The third complication introduced by disorder is that the topological invariant $q$ depends not only on the mean value of disorder strength but on its particular realization.
As a result, for a given wire length $L$ and for a fixed mean value of disorder strength, the topological number becomes a statistically distributed variable \cite{BagretsKamenev}.
Its mean $\corr{q}$ gradually varies from $-1$ to $1$ by changing the control parameter $\mu$, which drives the transition from the topological to trivial phase. 
It is only in the thermodynamic limit, $L \to \infty$, when the transition
between these phases becomes a sharp quantum phase transition. This happens due
to complete Anderson localization of both topologically trivial and nontrivial
insulating states on the two sides of the transition.

Direct experimental observation of quasiparticle transport in quantum wires of class D can be problematic due to
shunting effects of the superconducting condensate.
A possible way to address quasiparticle dynamics is by measuring \emph{thermal} rather than electrical transport properties \cite{ReadGreen_2000}. The thermal conductance $G$ of a mesoscopic system can be conveniently expressed in units of thermal conductance quantum $G_0 = \pi k_B^2 T / 6 \hbar$. The corresponding dimensionless conductance $g=G / G_0$ is then given by the standard Landauer formula as a sum of transmission probabilities: $g = \sum_n T_n$. Influence of potential disorder on the thermal conductance in Majorana wires was studied in a number of theoretical works \cite{Brouwer_delocalisation, GruzbergReadVishveshwara, Akhmerov_classD, BagretsKamenev, Eslam_thesis}, both analytically and numerically.

\begin{figure}
	\includegraphics[width=\linewidth]{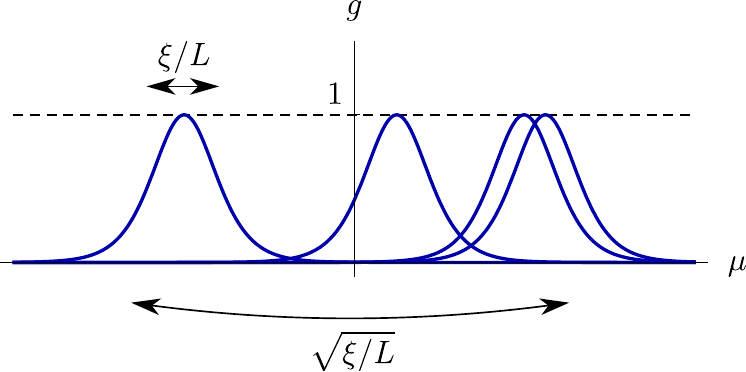}
	\protect\caption{Dependence of the dimensionless conductance $g$ of a wire of length $L\gg\xi$ on a control parameter $\mu$ for a number of disorder realizations (sketch). Conductance reaches its maximal value $g=1$ at $\mu^* = - v \lambda' / L$ (see discussion in the text). The peak has a width that scales as $1/L$, while peak centers are distributed with a typical width scaling as $1/\sqrt{L}$.
}
\label{fig:peaks}
\end{figure}

The simplest description of quasiparticle transport in disordered quantum wires of class D in the vicinity of the topological transition is achieved for a single propagating transverse mode (\emph{one-channel case}) \cite{SheltonTsvelik, BalentsFisher, Akhmerov_classD}. Its low-energy physics is governed by a one-dimensional (1D) random-mass Dirac Hamiltonian $\mathcal{H} = -i v \sigma_z \partial / \partial x + [\mu + m(x)] \sigma_y$, where $\sigma_i$ are Pauli matrices, $\mu$ is a control parameter and a position-dependent $m(x)$ fluctuates around zero mean (we assume it to be short-correlated).
For this model, the Lyapunov exponent $\lambda$, which determines the reflection coefficient $r = \tanh \lambda$ and dimensionless conductance $g = T = 1/\cosh^2 \lambda$, undergoes a drifted random walk with the increase of the wire length $L$: $\lambda = \mu L/v + \lambda'$, where $\lambda' = \int_0^L dx \, m(x) / v$.
Due to the central limit theorem, $\lambda'$ is a normally-distributed random variable that fluctuates around zero with the variance $\var \lambda' = L / \xi$, where $\xi = 2 l$ is the
disorder-dependent
\textit{correlation length}, $l$ is elastic mean free path and
the factor of 2
accounts for the presence of two counterpropagating modes.
In an infinite system, the fluctuating component $\lambda'$ becomes irrelevant, and the transition between the topological phases takes place right at $\mu=0$. For a finite system, the topological transition becomes smeared, see Fig.~\ref{fig:peaks}. For a given disorder realization $m(x)$, ideal transmission takes place at $\mu_* = - v\lambda'/L$, with the conductance decaying from unity at $\mu-\mu_*\sim v/L$. Sample-to-sample fluctuations of $\mu_*$ are characterized by $\var\mu_* = v^2 / L \xi$.

Below we will be mainly interested in the \emph{critical regime} realized for $\mu=0$, when the drift term for the Lyapunov exponent vanishes and the \textit{localization length} $\xi_\text{loc} =  v / |\mu|$  diverges.
In this case, random walk for $\lambda$ results in the zero-centered normal distribution
$P(\lambda)$,
which in the limit $L\gg\xi$ translates to the following distribution of the transmission coefficient $T$ (and so of $g$) \cite{Beenaker_Dorokhov}:
\be
\label{one_channel_Dorokhov}
	P(T) = \frac{\corr{g}}{2} \frac{1}{T \sqrt{1 - T}} ,
\ee
where $\gavg$ is defined by the value of
$P(\lambda=0)$:
\be
\label{g-DMPK}
  \langle g \rangle
  = \frac{2}{\sqrt{2 \pi \var \lambda}} =
  \sqrt{\frac{2 \xi}{\pi L}}.
\ee
Equation \eqref{one_channel_Dorokhov} should be corrected at smallest $T\sim e^{-2/\corr{g}}$ due to roll-off of $P(\lambda)$.
Remarkably, the distribution~\eqref{one_channel_Dorokhov}
formally coincides with the Dorokhov bimodal distribution \cite{Dorokhov},
which is known to describe transmission eigenvalues \textit{density} for multi-channel disordered wires in the Drude regime, where conductance is a self-averaging quantity. Contrary to that, Eq.\ \eqref{one_channel_Dorokhov} refers to the single-channel case, when conductance strongly fluctuates on the interval $0 < g < 1$,
with all its moments $\corr{g^n}$ sharing the scaling of $\corr{g}\sim1/\sqrt{L}$. In particular, for the variance $\var g=\corr{g^2}-\corr{g}^2$ one finds
\be
\label{var-g}
  \var g = \frac{2}{3} \corr{g} .
\ee
The absence of self-averaging can be also seen by comparing the scaling of $\corr{g}$ with the stretch-exponential decay of the typical conductance,
$g_\text{typ} = \exp \langle \log g \rangle \sim \exp(-4\sqrt{L / 2\pi\xi})$.

Note that at the critical point considered, half of disorder realizations belongs to the trivial phase and another half belongs to the topological phase.
At large $L$, the majority of realizations have an small conductance, $g\ll1$. However for relatively rare configurations (probability decreases as $1/\sqrt{L}$) the sample is close to the phase transition with $g\sim 1$, and these very configurations determine the average conductance $\langle g \rangle \sim 1/\sqrt{L}$, as well as all its higher moments.
To understand the scaling $\gavg \propto 1 / \sqrt{L}$, note that in the critical regime ($\mu = 0$) the probability to get such near-critical configuration in the process of disorder sampling can be assessed as the ratio of the single peak width $\sim1/L$ to the width of the peak centers distribution $\sim 1/\sqrt{L}$ (see Fig.~\ref{fig:peaks}).
Tuning the control parameter $\mu$ away from the transition puts all realizations to the same topological phase, with an exponentially small conductance in the long-wire limit.
Similar outcomes were obtained in Ref.~\onlinecite{Motrunich} with the help of real-space strong-disorder renormalization group and transfer matrix approaches.

In the \textit{multi-channel case}, localization in quantum wires of class D was considered in Refs.\ \cite{Brouwer_delocalisation, GruzbergReadVishveshwara} within the Dorokhov-Mello-Pereyra-Kumar (DMPK) approach \cite{Dorokhov, MelloPereyraKumar}. Solving the Fokker-Planck  equation for the distribution function of transmission eigenvalues $T_n=1/\cosh^2 \lambda_n$ \emph{in the long-wire limit} leads to the usual ``crystallization'' of the Lyapunov exponents $\lambda_n$ \cite{EversMirlin}.
However, the peculiarity of class D fine-tuned to the critical point
(the opposite case requires a special treatment \cite{GruzbergReadVishveshwara})
is that the lowest exponent $\lambda_1$ is distributed normally near zero (ideal transmission) with the variance
$\var \lambda_1 = L /  \xi$,
where $\xi = 2 N l$ is the correlation length of the $N$-channel wire (compare with the $N=1$ case considered above). With the contribution of other channels being exponentially suppressed, the conductance $g=T_1$ is determined by the most transparent channel, making this regime completely analogous to the one-channel model discussed above.
Hence, in the limit $L\gg\xi$, Eqs.\  \eqref{one_channel_Dorokhov} and \eqref{g-DMPK} are applicable for multichannel wires as well, with $T = T_1$. Such a behaviour is also observed in other topological classes
tuned to a critical regime in 1D, a phenomenon called ``superuniversality'' in Ref.~\onlinecite{GruzbergReadVishveshwara}.

A weak point of the DMPK approach is that it allows to calculate transport properties in the crossover between the Drude and localization regimes only in special cases.
This happens when the underlying Fokker-Planck equation can be solved with the help of the Sutherland transformation: in 
class A \cite{BeenakkerRejaei_1993, Frahm_1995, Macedo_2006}, CI and DIII \cite{Brouwer_delocalisation, Macedo_2018}, and AIII \cite{Mudry_Brouwer_Furusaki}. 

In order to trace the dependence of conductance moments on the wire length $L$ for class D, one should resort to a complimentary sigma-model technique \cite{Efetov-book}. The average conductance $\corr{g(L)}$ was calculated in this way in Refs.\ \cite{BagretsKamenev, Eslam_thesis}
(the difference from our result \eqref{result_general} by an overall numerical factor is due to an apparently different normalization of the dimensionless conductance
\cite{com-factors}).
A striking feature of the symmetry class D is that the target space of the supersymmetric sigma model consists of two connected components \cite{Zirnbauer1996,BocquetZirnbauer}.
Remarkably, localization happens only if one allows for jumps (domain walls) between the two components \cite{BocquetZirnbauer, ReadLudwig2000}. Such processes are described by an additional term in the sigma-model action proportional to $\log \tilde\chi$, where $\tilde\chi$ is the so-called kink fugacity \cite{BagretsKamenev}, characterizing the deviation from the  critical regime and thus resulting in a finite localization length $\xiloc = 2 \xi / \tilde{\chi}^2$  \cite{BagretsKamenev}.

\begin{figure}
	\includegraphics[width=\linewidth]{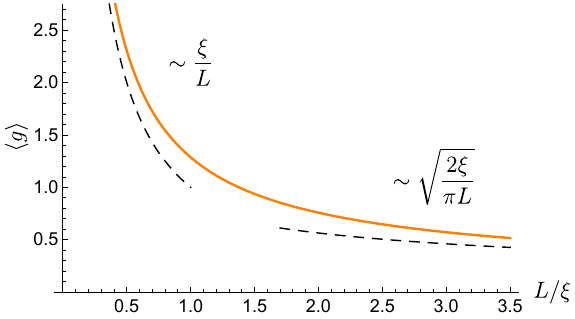}
	\protect\caption{Disorder-averaged conductance of quantum wires of class D at criticality as a function of the wire length $L$. Dashed lines show the leading short- and long-wire asymptotics.}
\label{fig:conductance}
\end{figure}

In the following we will focus exclusively on the \textit{critical} regime with completely suppressed kinks ($\tilde{\chi} = \mu = 0$). This regime corresponds to the DMPK approach of Ref.~\onlinecite{Brouwer_delocalisation} discussed above, which is characterised by an algebraic decay of the average conductance at $L\gg \xi$ given by Eq.\ \eqref{g-DMPK} and a rather involved analysis is required to see localization in the DMPK framework \cite{GruzbergReadVishveshwara}.
The critical regime can be defined in terms of physical quantities at arbitrary $L$ as the regime when the average determinant of the reflection matrix is zero: $\langle \det r \rangle = 0$.
The dependence of $\corr{g(L)}$ for class D at criticality obtained in Ref.~\onlinecite{BagretsKamenev} is shown in Fig.\ \ref{fig:conductance}. It smoothly interpolates between the Drude regime with $\corr{g}=\xi/L$ to the long-wire regime with $\corr{g}$ given by Eq.\ \eqref{g-DMPK}.

In the present paper we make a next step in the analysis of quasiparticle transport in multi-channel quantum wires of class D at criticality and derive exact expressions for the conductance variance $\var g$, its third cumulant, average noise power, and $\var\det r$ at arbitrary $L/\xi$ in the diffusive regime.
Calculation of all those quantities requires averaging of four Green's functions that cannot be done within the one-replica
supersymmetric
sigma model considered in Refs.\ \cite{BagretsKamenev, Eslam_thesis}, forcing us to consider a two-replica ($n=2$) version of the supersymmetric sigma model.
The heat kernel for the latter is obtained with the use of the Iwasawa decomposition trick \cite{Zirnbauer1991, Zirnbauer1992, MMZ}.

The main technical achievement of this paper is classification of radial eigenfunction of a higher-rank
(several replicas)
supersymmetric sigma model. We find that the straightforward approach for their construction outlined in Refs.\ \cite{Zirnbauer1991, Zirnbauer1992, MMZ} produces an incomplete basis due to vanishing of Grassmann integration of ``too symmetric'' Iwasawa wave functions. This problem is solved by including additional subfamilies of eigenfunctions with a smaller amount of quantum numbers, which are intimately related to the radial eigenfunctions of the sigma model with a smaller number of replicas. This finding is expected to be relevant for the heat kernel construction for all higher-rank ($n>1$) sigma models of arbitrary symmetry classes.

The paper is organized as follows.
In Sec.\ \ref{S:sum-res} we summarize new physical results obtained in this work.
In Sec.\ \ref{S:mathpre} we introduce the main mathematical ingredients required for construction of the heat kernel of the supersymmetric sigma model. The outlined procedure for class D with two replicas is implemented in Sec.\ \ref{sec:eigenfunctions_constructions}, where we introduce an additional subfamily of radial eigenfunctions and discuss the behavior of eigenfunctions at particular lines needed to extract their normalization and behavior at the origin. The final expressions for the conductance variance, its third cumulant, shot-noise power and average square of the determinant of the reflection amplitude matrix are obtained in Sec. \ref{sec:results}. The technique developed and results obtained are discussed in Sec.\ \ref{sec:conclusion}. Important technical details are relegated to numerous Appendices.

\begin{figure}
	\includegraphics[width=\linewidth]{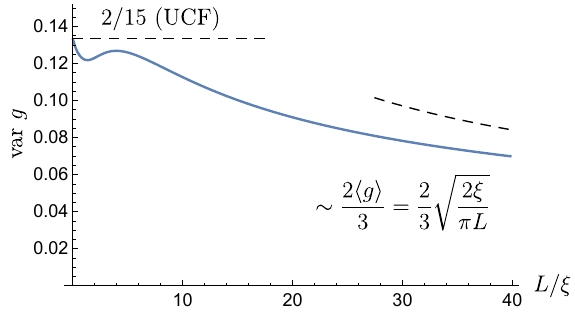}
	\protect\caption{Dependence of the conductance variance for class D at criticality on the wire length $L$. Dashed lines show the short- and long-wire asymptotics.
}
\label{fig:variance}
\end{figure}

\section{Summary of results}
\label{S:sum-res}

\begin{figure}
	\includegraphics[width=\linewidth]{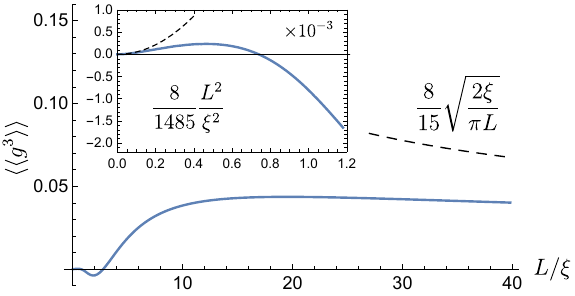}
	\protect\caption{Dependence of the third cumulant
$\corr{\corr{g^3}}$
of the conductance for class D at criticality on the wire length $L$. Inset: short-wire part of the dependence. Dashed lines show the short- and long-wire asymptotics.
}
\label{fig:cumGGG}
\end{figure}

\begin{figure}
	\includegraphics[width=\linewidth]{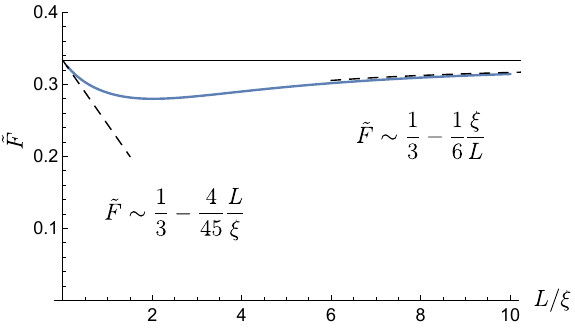}
	\protect\caption{Length dependence of the pseudo-Fano factor \eqref{pseudo-Fano}, which determines the average thermal shot-noise power of the quasiparticle current.
Dashed lines show the short- and long-wire asymptotics.}
\label{fig:QFano}
\end{figure}

\begin{figure}
	\includegraphics[width=\linewidth]{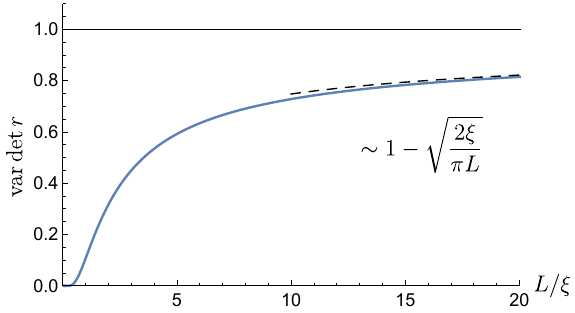}
	\protect\caption{Length dependence of the average square of the determinant of the reflexion matrix, $\langle (\det r)^2 \rangle = \var \det r$.
The fact that the curve approaches the asymptotic value of $1$ indicates that for every sample $\det r = \pm 1$ in the limit $L \rightarrow \infty$.
}
\label{fig:ChiChi}
\end{figure}

The results of our study are presented graphically in Figs.~\ref{fig:variance}--\ref{fig:ChiChi}. Figure~\ref{fig:variance} depicts the dependence of the conductance variance on the wire length, illustrating the crossover behavior from the value of $\var g = 2/15$ at $L\ll\xi$ in the Drude regime
(universal conductance fluctuations (UCF) \cite{UCF_Lee_Stone}) to the ``superuniversal'' limit $\var g = (2/3)\gavg$ at $L \gg \xi$, as given by Eq.\ \eqref{var-g}.
Contrary to a featureless dependence of $\corr{g}$ with the crossover around $L/\xi\sim1$ (see Fig.\ \ref{fig:conductance}), the dependence of $\var g$ on $L$ exhibits a reentrant behavior, with the crossover being strongly displaced towards larger wire lengths $L/\xi\sim15$.

Figure \ref{fig:cumGGG} depicts the dependence of the third cumulant of conductance, $\corr{\corr{g^3}} = \langle g^3 \rangle - 3 \langle g^2 \rangle \langle g \rangle + 2 \langle g \rangle^3$, on the wire length. This quantity determines the asymmetry of the conductance distribution (skewness) about its average. In the quasiclassical limit $L \ll \xi$, the third cumulant is $\corr{\corr{g^3}} \sim (L/\xi)^2$ with a very small numerical coefficient. At longer distances it changes sign twice before approaching the asymptotic dependence $\sim \sqrt{\xi/L}$ at $L \gtrsim 40$.

The two-replica sigma-model also allows one to calculate the \textit{thermal} shot noise power, which corresponds to counting quasiparticles irrespective of their charge and differs from a usual \textit{electrical} shot noise power. 
We will characterise the former by the pseudo-Fano factor 
\be
\label{pseudo-Fano}
\tilde F= \frac{\big\langle \sum_n T_n (1 - T_n) \big\rangle }{ \big\langle \sum_n T_n \big\rangle },
\ee
which is determined by the ratio of the average transport moments instead of the average ratio.
The length dependence of $\tilde F$ is shown in Fig.~\ref{fig:QFano}.
Remarkably, it equals $1/3$ both in short- and long-wire limits, as both of them are characterised by the Dorokhov distribution \eqref{one_channel_Dorokhov}. For short wires, the situation is typical for diffusive metals, where Dorokhov distribution describes transmission eigenvalues density and Fano factor is a self-averaging quantity due to the aggregated contribution of many channels. Contrary to that, for long wires, transport is provided by only one mode with the lowest Lyapunov exponent. Since it is  described by the same Eq.\ \eqref{one_channel_Dorokhov}, we arrive at the same value of $\tilde F=1/3$, with the actual Fano factor $F$ exhibiting strong sample-to-sample fluctuations.

Yet another way to characterize topological properties of the wire is to study moments of the determinant of the reflection amplitudes, $\det r$. Though different from the true topological number $q = \sign \det r$, this quantity shows similar behaviour of interpolating between $\pm 1$ as the control parameter is driven across the phase transition. Its average value $\corr{\det r}$ as a function of $L$ and kink fugacity $\tilde\chi$ was calculated in Ref.~\onlinecite{BagretsKamenev}. Vanishing right at the critical line $\tilde\chi=0$, $\corr{\det r}$ flows to $\pm1$ with increasing $L$ for any finite bare kink fugacity $\tilde\chi$, in a sense similar to the renormalization-group flow of $\sigma_{xy}$ in the integer quantum Hall effect \cite{Khmel,Pruisken}.

Here we calculate the second moment
of the determinant,
$\corr{\det^2r}$, and demonstrate that it does not vanish at the critical line (where $\corr{\det r}=0$). With the increase of the wire length $L$, it interpolates between 0 (short wires) and 1 (long wires), as shown Fig.\ \ref{fig:ChiChi}. Taking into account that $|\det r| \leq 1$ due to unitarity of quantum mechanics, one concludes that in the limit $L \rightarrow \infty$, $\det r$ takes values $+1$ and $-1$ for any specific sample with equal probabilities. That
illustrates our statement about the large-$L$ behaviour from a sigma-model perspective: even in the critical regime, almost all of the samples are insulating and topological/trivial in equal proportion.

\section{Mathematical preliminaries}
\label{S:mathpre}

In this preparatory Section, we outline the main steps towards calculation of $\corr{g}$, $\corr{g^2}$, and $\corr{g^3}$ in the sigma-model formalism and introduce necessary mathematical concepts. Our analysis closely follows a pathway developed in Refs.\ \cite{Zirnbauer1991, Zirnbauer1992, MMZ} for conventional symmetry classes and implemented for calculation of $\corr{g}$ in quantum wires of class D in Refs.\ \cite{BagretsKamenev, Eslam_thesis}. The peculiarity of class D is that calculation of $\corr{g^2}$ already requires the use of the sigma model with two replicas ($n=2$), which significantly complicates the whole analysis.

In this Section, we will try to balance between generality and concreteness.
On the one hand, we will keep our discussion as general as possible, without resorting to a particular basis (that will be done later in Sec.\ \ref{sec:eigenfunctions_constructions}). Thus almost all formulas of this Section could be easily applied to other symmetry classes with an arbitrary number of replicas. On the other hand, some formulas below do rely on the particular symmetry class D with $n=2$, both to illustrate the general formalism and to prepare ingredients for actual calculations in the next Sections.

\subsection{Sigma model and conductance moments}
\label{sec:sigma-model definition}

The sigma-model action for the $N$-channel quantum wire of class D in the critical regime has the usual form
($\xi = 2 N l$ is the correlation length)
\cite{BagretsKamenev, Eslam_thesis, com-factors}:
\be
\label{sigma_model_action}
	S[Q] = - \frac{\xi}{16} \int_{0}^{L}dx\, \str (\nabla Q)^{2} ,
\ee
where $Q \in \text{BF} \otimes \text{N} \otimes \text{R}$ is a supermatrix, which lies in the tensor product of the Bose-Fermi (BF), Nambu-Gor'kov (N) and replica (R) spaces, and is subject to the charge conjugation constraint $\overline{Q} = C^T Q^T C = - Q$ due to the particle-hole structure of the BdG Hamiltonian. The matrix $C$ is orthogonal, $C C^T = \mathbb{1}$, and satisfies $C^2 = - k$, where $k$ distinguishes bosons and fermions and has the following structure in the BF space: $k = \{\mathbb{1}, - \mathbb{1}\}_\text{BF}$.
A special role in the theory is played by a selected matrix $\Lambda$ (origin), satisfying $\overline\Lambda = - \Lambda$, $\str \Lambda = 0$ and $\Lambda^2 = \mathbb{1}$. The whole sigma-model manifold can be obtained by rotating $\Lambda$ with elements $T$ of a certain supergroup $G$:
\be
\label{Q_via_Lambda_and_T}
	Q = T^{-1} \Lambda T,
\ee
where $\overline{T} =  T^{-1}$.
Then the sigma-model manifold is a coset (\textit{symmetric space}) $G / K $, where $K$ is the subgroup in $G$ that commutes with $\Lambda$: $[K, \Lambda] = 0$.

For class D with $n$ replicas,
$G$ is the supergroup $\text{SpO}(n,n | 2n)$,
$K$ is the supergroup $\text{U}(n | n)$, so that the sigma-model manifold is the coset $\text{SpO}(n,n | 2n) / \text{U}(n|n)$.
It generally consists of two disconnected submanifolds. For example, the supermanifold $\text{SpO}(1,1 | 2) / \text{U}(1|1)$ in the $n=1$ case is a hyperboloid $H_2$ as its Bose-Bose (BB) sector and a set of two points ($\mathds{Z}_2$) in its Fermi-Fermi (FF) sector \cite{BagretsKamenev, Eslam_thesis}. The critical point between the topological and trivial phases that we address in this paper corresponds to the absence of jumps between the disconnected components of the sigma-model manifolds (kinks) \cite{BocquetZirnbauer, ReadLudwig2000}, that allows us to consider only one connected component,
namely $\text{SpSO}(1,1 | 2) / \text{U}(1|1)$.
Away from the critical point, proliferation of kinks described by an additional term in the sigma-model action \eqref{sigma_model_action} leads to exponential localization both in the topological and trivial phases \cite{BagretsKamenev}.

While the averaged conductance $\corr{g}$ can be calculated from $n=1$ sigma-model,
evaluation of its higher moments generally requires higher $n$. Nevertheless for most classes, $\corr{g^2}$ (and hence $\var g$) can be calculated already from $n=1$ sigma-model, as in the supersymmetric approach two copies of the system (bosons and fermions) are averaged over disorder, each suitable for calculation of conductance in the noninteracting case \cite{MMZ}. However, the peculiarity of class D is that its FF sector in $n = 1$ case is empty (just two disconnected points) and therefore cannot be used to access the second copy of the system.

Thus for calculation of $\corr{g^2}$ and $\var g$ in class D one inevitably has to use the sigma model \emph{with two replicas} ($n=2$). Its supermanifold is $\text{SpO} (2,2 | 4) / \text{U}(2|2)$, with the BB sector being a rank-$2$ symmetric space $\text{Sp}(2, 2) / \text{U}(2)$ and the FF sector being a rank-$1$ symmetric space $O(4) / \text{U}(2) \simeq S_2 \times \mathds{Z}_2$ \cite{Helgason}, where the latter is isomorphic to the union of two disjoint spheres.

\color{black}

In the sigma-model language,
the moments of conductance
can be calculated by taking the derivative of the partition function with respect to an infinitesimal twist of the boundary conditions \cite{LL1993, LLYa1995, MMZ, KSO2016, Eslam_thesis}. The partition function is defined as a functional integral
\be
\label{partition_function}
  Z[\theta_i]
  =
  \int_{Q(0) = \Lambda}^{Q(L) = Q_L} \mathcal{D}[Q(x)] e^{-S[Q]} ,
\ee
where $Q_L = \Lambda \exp(\check{\theta})$ and the matrix of Cartan angles $\check\theta$ is defined in Eq.\ \eqref{theta-def} below.

The number of Cartan parameters $\theta_i$ depends on the number of replicas. Let $N_B$ ($N_F$) be the number of generators of Cartan algebra from the BB (FF) sector.
For class D with $n$ replicas,
\be
\label{N_B_and_N_F}
	N_B = n, \qquad N_F = \left\lfloor{n/2}\right\rfloor ,
\ee
where the floor brackets $\left\lfloor \cdot \right\rfloor$ denote the integer part of a number  (for even $n$, Eq.\ \eqref{N_B_and_N_F} was derived in  Ref.~\onlinecite{GruzbergMirlinZirnbauer_classification}).

Knowledge of $Z[\theta_i]$ allows us to compute a number of average physical quantities, see Appendix \ref{app:transport_via_SM}. The first three conductance moments can be expressed as follows:%
\begin{subequations}
\label{G_through_Z}
\begin{gather}
\label{G1_through_Z}
	\left\langle g \right\rangle = -4 \left.\frac{\partial^{2}Z(\theta_i)}{ \partial \thone^2}\right|_{0},
\\{}
\label{G2_through_Z}
	\left\langle g^2 \right\rangle = 16 \left.\frac{\partial^{4}Z(\theta_i)}{ \partial \thone^2 \partial \thtwo^2}\right|_{0} ,
\\{}
\label{G3_through_Z}
	\left\langle g^3 \right\rangle = -32 \left.\frac{\partial^{6}Z(\theta_i)}{ \partial \thone^2 \partial \thtwo^2 \partial \thF^2}\right|_{0} , \end{gather}
\end{subequations}
where the derivatives must be computed at the origin, $\theta_i=0$ [higher moments would require a sigma-model with a larger $n \geq 3$ number of replicas and thus more Cartan angles $\theta_i$].
The pseudo-Fano factor \eqref{pseudo-Fano} is given by the following expression:
\be
\label{F_through_Z}
	\tilde{F} = \frac{1}{3} + \frac{1}{9} \left. \left( \frac{\partial^2 Z}{\partial \thone^2} \right)^{-1} \left[ 4 \frac{\partial^4 Z}{\partial \thone^4} - \frac{\partial^4 Z}{\partial \thF^4} \right] \right|_{0} .
\ee
Finally, the average squared determinant of the matrix of reflection coefficients equals the partition function in the so-called ``south pole'' point. As mentioned in Introduction, in the critical regime $\langle \det r \rangle = 0$, so that
\be
\label{var_det_through_Z}
	\var \det r = \langle \text{det}^2  r \rangle = Z(\thone = 0, \thtwo = 0, \thF = \pi) .
\ee

In order to calculate the averages listed above 
at arbitrary wire length, we will need two different parametrizations of the $Q$-matrix manifold: Cartan parametrization, which
explicitly enters Eq.\ \eqref{G_through_Z}, and Iwasawa parametrization, which possesses the simplest form of the radial Laplace-Beltrami operator.

\subsection{Cartan-Efetov parametrization}
\label{sec:Cartan_paramertization}

Cartan parametrization (also referred to as Efetov parametrization in the sigma-model context) is obtained by applying Cartan decompostion to the $T$ matrix with respect to the involution $T \rightarrow \Lambda T \Lambda$. That allows to decompose $T = U_1 e^{\check{\theta} / 2} U$, where $U_1$ and $U$ commute with $\Lambda$ ($U\in K$, $U_1 \in K$) and $\check{\theta}$ lies in the
maximal abelian (Cartan)
subalgebra of matrices from $G$ that anticommute with $\Lambda$: $\{\Lambda, \check{\theta}\}$ = 0. Such a parametrization is redundant, so we choose $U_1$ to run over the whole group $K$ and leave in $U$ only the necessary number of parameters. The  $Q$ matrix does not depend on $U_1$ and acquires the form:
\be
\label{Cartan_parametrization}
  Q = U^{-1} \Lambda e^{\check{\theta} } U.
\ee

According to Eq.\ \eqref{N_B_and_N_F}, Cartan algebra of $n = 1$ sigma model  is parametrised by one parameter ($\theta_B$), originating from the BB sector, while for the two-replica case ($n=2$), $\check{\theta}$ can be represented as a linear combination of three commuting generators $h_i$, two from the BB sector and one from the FF sector:
\be
\label{theta-def}
  \check{\theta}
  =
  \thone \check{h}_{B1} + \thtwo \check{h}_{B2} + i \thF \check{h}_{F},
\ee
where $\theta_i$ are real on the sigma-model manifold.

An important mathematical structure is the \textit{root system} with respect to $h_i$. It consists of matrices $Z_\alpha$, called \textit{root vectors}, which are eigenvectors for all $h_i$ acting in the adjoint representation: $[\check{\theta}, Z_\alpha] = \alpha(\theta) Z_\alpha$, where $\alpha$ is a linear function on the Cartan algebra, called a \textit{root}. \emph{Positive roots}\ are chosen as a subset $\mathcal{R}_+$ that lies in a selected half-plane in the dual vector space.
Peculiarity of supermanifolds with respect to the well-known noncompact symmetric spaces \cite{Helgason} is the fact that root vectors belonging to off-diagonal blocks in the BF space are Grassmann numbers. The corresponding roots should be counted with negative multiplicities \cite{Zirnbauer1991,MMZ}. The root system of class D with $n=2$ is presented in Table \ref{T:roots} and Fig.\ \ref{fig:roots}, see Appendix~\ref{app:basis_and_root_system}.

The root system is symmetric with respect to the so-called Weyl group, which is generated by reflections with respect to the planes perpendicular to each
commuting
root and thus acting in the dual Cartan space.
Studying the action of the same group on the original Cartan space allows one to choose its minimal domain, which is called a \textit{Weyl chamber}.
The values of the Cartan angles $\check\theta$ in parametrization \eqref{Cartan_parametrization} should be restricted to this domain. Otherwise, the parametrization would be redundant, as the whole Cartan space is already covered by the $U$ matrix [see \eqref{Cartan_parametrization}].

In the following, we will consider radial wavefunctions that depend only on $\theta_i$ and obey the symmetries given by the Weyl group. 
In the supersymmetric case, the Weyl group consists of the BB and FF sectors. 
In class D with two replicas, FF part of the Weyl group ensures radial wavefunctions are even functions of $\thF$, while BB part is generated by sign flips of $\thone$, $\thtwo$ and their interchange ($\thone \leftrightarrow \thtwo$).
That's why we will study radial functions only in one Weyl chamber, which we choose to be $\thone \geq \thtwo \geq 0$.

The measure for Cartan parametrization \eqref{Cartan_parametrization} can be written as $DQ = J \, DU \, D\theta$.
A beneficial property of this
parametrization is the factorization of the Jacobian $J = J_U J(\theta_i)$ into the Haar measure $J_{U}$ on the group $K$ and $\theta$-dependent part $J(\theta_i)$ \cite{Helgason, Eslam_thesis}. The latter can be explicitly expressed as a product of factors corresponding to each positive root $\alpha \in R^+$:
\be
\label{J_through_roots}
 	J(\theta_i) =  \prod_{\alpha \in R^+} \left[ \sinh \alpha(\theta) / 2 \right]^{m_\alpha},
\ee
where $m_\alpha$ are roots multiplicities. Applied to class D with two replicas this formula yields:
\be
\label{Jacobian}
  J(\theta_i)
  = \frac{(\cosh \thone - \cosh \thtwo) \sinh \thone \sinh \thtwo \sin \thF}
	{(\cosh \thone - \cos \thF)^2 (\cosh \thtwo - \cos \thF)^2} .
\ee

Cartan parametrization is especially important since according to Eq.\ \eqref{G_through_Z} it is directly related to the averaged conductance and its higher moments. Therefore our main goal will be to calculate the partition function $Z[\theta_i]$.

\subsection{Transfer-matrix Hamiltonian and the heat kernel}

A standard method for evaluating the partition function \eqref{partition_function} for 1D systems is switching from the functional integral representation to the Schr{\"o}dinger-like equation for the wavefunction $\psi(Q, t)$ \cite{EL1983}. Its evolution is governed by the so-called transfer-matrix Hamiltonian, with the spatial coordinate $x$ playing role of imaginary time:
\be
\label{Schroedinger_equation}
	\frac{\xi}{2} \partial_x \psi (Q, x) = - \hat{H} \psi(Q, x) .
\ee
The Hamiltonian is given by the Laplace-Beltrami operator on the sigma-model target space:
\be
\label{Hamiltonian_is_Laplacian}
	\hat{H} = - \Delta =  - \frac{1}{J} \partial_\alpha J \mathfrak{g}^{\alpha \beta} \partial_\beta ,
\ee
where $\mathfrak{g}_{\alpha \beta}$ is the metrics induced by the expression $dl^2 = (-1/2) \str dQ^2 = g_{\alpha \beta} dX^{\alpha} dX^{\beta}$, where $X^{\alpha}$ are the coordinates on the sigma-model supermanifold \cite{Efetov-book, Eslam_thesis}.
Its diagonalization is provided by the set of eigenfunctions $\phi_\nu(Q)$ satisfying
\be
\label{eig}
  \Delta \phi_\nu(Q) = - \epsilon_\nu \phi_\nu(Q) .
\ee

The partition function \eqref{partition_function} 
coincides with the heat kernel for the Schr{\"o}dinger equation \eqref{Schroedinger_equation}:
\be
\label{Z_eq_psi}
	Z[\theta_i] = \psi(Q_L, x=L) ,
\ee
which is obtained by solving it with the initial condition $\psi(Q, x=0) = \delta(Q, \Lambda)$, where $\delta(Q, \Lambda)$ is a supersymmetric delta-function, which equals to 1 at the origin \cite{Efetov-book}:
\be
\label{supersymmetric_delta_function}
	\delta(Q, \Lambda) =
	\begin{cases}
		1, \quad Q = \Lambda, \\
		0, \quad Q \neq \Lambda .
	\end{cases}
\ee

Knowledge of the eigenfunctions allows one to write down the spectral representation of the heat kernel:
\be
\label{heat_kernel}
	\psi(Q,x) = \sum_\nu \mu_\nu \phi_\nu(Q) e^{-2 \epsilon_\nu x / \xi},
\ee
where summation also includes integration over continuous quantum numbers and $\mu_\nu$ are the coefficients of $\delta(Q,\Lambda)$ in the basis $\phi_\nu(Q)$ (see Sec.~\ref{SS:measure}).
As both the Hamiltonian and the initial condition are invariant with respect to rotations by $U\in K$,
so is the heat kernel: $\psi(U^{-1}QU, x) = \psi(Q, x)$.
Therefore only \textit{radial} eigenfunctions of the Laplace operator (the ones that depend on Cartan angles $\theta_i$ only) enter expression \eqref{heat_kernel},
that greatly simplifies the analysis. For this reason instead of the full Laplacian \eqref{Hamiltonian_is_Laplacian} we will need only its part that acts on $\theta_i$ variables. This part, called the \textit{radial Laplacian}, can be obtained by taking $\theta_i$-block of $\mathfrak{g}_{\alpha \beta}$, which we denote as $g_{i j}$ (see Appendix \ref{app:basis_and_root_system}):
\be
\label{g_metrics}
	dl_{\text{rad}}^2 = \sum_{ij} g_{i j} \theta_i \theta_j  = \thone^2 + \thtwo^2 + 2 \thF^2 .
\ee
Then from Eq.\ \eqref{Hamiltonian_is_Laplacian} with $\mathfrak{g}^{\alpha \beta}$ replaced by $g^{ij}$ we get the following expression for the radial Laplacian:
\be
\label{radial_laplacian}
	\Delta_{\text{rad}} = \frac{1}{J} \left( \frac{\partial}{\partial \thone} J \frac{\partial}{\partial \thone}
	+ \frac{\partial}{\partial \thtwo} J \frac{\partial}{\partial \thtwo}
	+ \frac{1}{2} \frac{\partial}{\partial \thF} J \frac{\partial}{\partial \thF}
	\right) .
\ee

\black

The expansion of the heat kernel in terms of the eigensystem of the Laplace-Beltrami operator given by Eq.\ \eqref{heat_kernel} is very generic.
However in the supersymmetric case,
one typically adds unity to the r.h.s.\ of Eq.\ \eqref{heat_kernel} \cite{Zirnbauer1992, MMZ, BagretsKamenev, Eslam_thesis}. Note that $\phi_0=1$ is just the zero mode of the Hamiltonian and therefore this spurious unity is already contained in the expansion \eqref{heat_kernel}.
The reason why it is added by hands is that the procedure of the eigenfunction construction implemented by many authors fails to reproduce the zero mode, which then should be restored manually.
However as the number of replicas grows and the sypersymmetric space becomes more complicated, the number of eigenfunctions that cannot be obtained by averaging the plane wave in the Iwasawa parametrization over the group $K$ also grows and one has to reconsider this issue. That will be done in Secs.\ \ref{sec:Iwasawa_parametrization} and \ref{SS:eigenfunctions}.

\subsection{Iwasawa parametrization}
\label{sec:Iwasawa_parametrization}

Fourier analysis for symmetric spaces has been developed by Harish-Chandra \cite{Helgason} and generalized to the sypersymmetric case by Zirnbauer \cite{Zirnbauer1992, MMZ}. To construct the eigenbasis of the Laplace operator it is convenient to resort to the so-called Iwasawa parametrization:
\be
  Q = N^{-1} \Lambda e^{\check{a}} N.
\ee
It is obtained by applying Iwasawa decomposition on $T$ matrix: $T = U_I e^{\check{a} / 2}  N$, where $U_I \in K$, $\check{a}$ lies in Cartan subalgebra
\be
\label{a-def}
  \check{a} = \aone \check{h}_{B1} + \atwo \check{h}_{B2} + \aF  \check{h}_{F}
\ee
[the generators $\check h$ are the same as in Eq.\ \eqref{theta-def}] and $N \in \mathcal{N}_+$ lies in the exponential of the subalgebra of \textit{positive} roots represented by nilpotent matrices.
Iwasawa parametrization is characterised by the Jacobian, which is an exponential of a linear function: $J_I(a_i) = e^{\rho(a)}$, where $\rho$ is the so-called Weyl vector, expressed as the half-sum of the positive roots $\alpha(a)$ weighted with their multiplicities:
\be
\label{Weyl_vector}
	\rho(a)  =  \frac{1}{2} \sum_{\alpha \in R^+} m_\alpha \alpha(a) .
\ee
Note that choosing a particular set of positive roots breaks the symmetry between the variables $\aone$ and $\atwo$. For the choice specified in Appendix \ref{app:basis_and_root_system} and shown in Fig.\ \ref{fig:roots}, $\rho(a) = \aF - \atwo$.

The Laplace operator in Iwasawa coordinates $(a, N)$ takes a very simple form:
\be
\label{Iwasawa_laplacian}
  \Delta
  =
  \sum_i \left[ \frac{\partial}{\partial a_i} \cdot \frac{\partial}{\partial a_i} + 2 \rho_i (a) \cdot \frac{\partial}{\partial a_i} \right] + \Delta_N,
\ee
where $\rho_i$ are the components of the Weyl vector, dot product is defined by radial metrics $g^{ij} = (g_{ij})^{-1}$ [see \eqref{g_metrics}] and $\Delta_N$ is the $N$-part of the Laplacian that nullifies all functions that depend only on $a$: $\Delta_N f(a) = 0$.

Therefore plane waves
\be
\label{plane-wave}
  e^{i p a} = e^{i \sum_i p_i a_i}
\ee
are the eigenfunctions of the radial Laplace operator:
\be
\label{eig-Iw}
	\Delta e^{i p a} = - \epsilon_p  e^{ipa}, \qquad \epsilon_p = p \cdot p - 2 i \rho \cdot p .
\ee
The radial eigenfunctions (\ref{plane-wave}) in the Iwasawa representaton are parametrised by three momenta $p_i$, corresponding to the three-dimensional Cartan algebra.

\subsection{From Iwasawa to Cartan: general route}zation

In order to convert plane waves \eqref{plane-wave} in the Iwasawa coordinates to radial wave functions in the Cartan coordinates, one has to make two steps. First it is necessary to obtain an explicit expression for $a(\theta, U)$, which can be done by equating $\Lambda Q$ in Iwasawa and Cartan parametrizations:
\be
\label{LambdaQ_two_parametrizations}
  \Lambda N^{-1} \Lambda e^{\check{a}} N
  =
  U^{-1} e^{\check{\theta} } U ,
\ee
and solving the resulting set of equations.
However, when expressed in terms of Cartan coordinates $(\theta, U)$, the wave functions \eqref{plane-wave} will gain an unwanted $U$-dependence. Therefore the second step in obtaining the radial eigenfunctions would be to perform \emph{isotropization}\ of $e^{i p a}$ over the group $K$:
\be
\label{wavefunction_K_integral}
  \phi_p (\theta)
  =
  \bigl< e^{i p a(\theta, U)} \bigr>_K
  \simeq
  \int_{U \in K} e^{i p a(\theta, U)} .
\ee

Due to the presence of Grassmann variables, the last part of Eq.\ \eqref{wavefunction_K_integral} should be understood symbolically: If the integrand does not depend on (some) Grassmann variables for a given momentum $p$ then integration over them should not be done to ensure a nonzero value of $\phi_p(\theta)$. The simplest example is the case $p=0$ corresponding to the wave function $\phi_{0} = 1$, which already does not depend on $U$ and therefore can be used as is. However if we formally integrate it over the group $K$ the result will be zero. This is the reason why this ``too symmetric'' wave function is usually added to Eq.\ \eqref{heat_kernel} by hands.

However, as we discuss below, for class D with two replicas there exist yet another family of such exceptional ``too symmetric'' wave functions with one rather than three momenta that should be treated separately. This is the reason we refer to the process of radial eigenfunction construction from Iwasawa plane waves as isotropization rather than just averaging over the group $K$.

Finally, we note that the overall normalization factor in Eq.\ \eqref{wavefunction_K_integral} is left unspecified. It will be determined later for each eigenfunction family separately, see Sec.~\ref{SS:eigenfunctions}.

\section{Radial eigenfunctions}
\label{sec:eigenfunctions_constructions}

The procedure of eigenfunction construction outlined in Sec.\ \ref{sec:Iwasawa_parametrization} is generic for any symmetric space. But its implementation for a particular symmetry class requires some art of choosing the most appropriate parametrization. Moreover, the supersymmetry is known \cite{MMZ} to introduce additional complexity, which as we demonstrate below grows with the number of replicas.

\subsection{Basis and parametrizations}
\label{SS:basis}

Now we specify a particular basis, which significantly simplifies further calculations. We arrange commuting and Grassmann variables according to the BF
grading
matrix $k = \diag \{1,1,-1,-1,-1,-1,1,1 \}$, acting as $\pm1$ on bosonic and fermionic variables, respectively. Following Ref.~ \onlinecite{Eslam_thesis}, we choose the matrix $\Lambda$ to be completely antidiagonal, see Eq.\ \eqref{Lambda-def}.
The charge conjugation matrix $C$ is given by Eq.\ \eqref{C-def}.
The root system is presented in Appendix~\ref{app:basis_and_root_system}.
In this basis, the generators $\check h_i$ of Cartan subalgebra are diagonal and the matrices \eqref{theta-def} and \eqref{a-def} take the form:
\begin{align}
\nonumber
	\check{\theta} &= \diag \{ \thone, \thtwo, i \thF, i \thF, -i \thF, -i \thF, -\thtwo, -\thone \} ,
\\
	\check{a} &= \diag \{ \aone, \atwo, \aF, \aF, - \aF, - \aF, -\atwo, -\aone \} .
\end{align}

The crucial advantage of the chosen basis is that it allows for a simple and constructive solution of Eq.\ \eqref{LambdaQ_two_parametrizations} for $a(\theta,U)$, relying on the fact that positive root vectors can be chosen to be strictly upper triangular matrices. Then $N$ and $ \Lambda N^{-1} \Lambda$ in Eq.\ \eqref{LambdaQ_two_parametrizations} become upper and lower triangular matrices, respectively, with unities on the main diagonal.
Hence the principal (super)minors of the l.h.s.\ of Eq.\ \eqref{LambdaQ_two_parametrizations} contain only $a$ variables that can be used to extract the required dependence $a(\theta,U)$. Since the first three elements of $\check a$ already contain all three $a_i$, it is sufficient to consider only first three principal submatrices of Eq.\ \eqref{LambdaQ_two_parametrizations}, leading to the set of relations:
\begin{gather}
	e^{\aone} = \bigl[ U^{-1} e^{\check{\theta} } U \bigr]_{11} , \nonumber \\
\label{equating_minors}
	e^{\aone + \atwo} = \det \bigl[ U^{-1} e^{\check{\theta} } U \bigr]_{\text{1--2},\text{1--2}} ,  \\
	e^{\aone + \atwo - \aF} = \sdet \bigl[ U^{-1} e^{\check{\theta} } U \bigr]_{\text{1--3},\text{1--3}} . \nonumber
\end{gather}

Successively applying Eqs.\ \eqref{equating_minors}, we obtain $e^{\aone}$, $e^{\atwo}$ and $e^{\aF}$. Then raising them to the powers $i p_\text{B1}$, $i p_\text{B2}$ and $i p_\text{F}$, respectively, and multiplying the resulting monomials we obtain the expression for Iwasawa plane wave $e^{ipa}$ in the Cartan coordinates.

Radial wave functions in Cartan coordinates $\theta$ should be obtained by isotropization of plane waves $\corr{e^{ipa(\theta,U)}}_K$ over matrices $U\in K$ according to Eq.\ \eqref{wavefunction_K_integral}. However, due to a large number of independent degrees of freedom that parametrise $U$, the resulting expression for $\phi_\nu(\theta)$ cannot be obtained in a closed form. Fortunately, for calculating the conductance moments \eqref{G_through_Z}, the full knowledge of radial functions is not needed.
Instead it is sufficient to determine (i)
their asymptotic behavior at large $\theta$, which controls
the normalization and hence the coefficients $\mu_\nu$ in the spectral decomposition of the heat kernel \eqref{heat_kernel}, and (ii) behaviour at small $\theta$, which is needed to compute derivatives in Eqs.~\eqref{G_through_Z}--\eqref{var_det_through_Z}.

To make analytical extraction of the large-$\theta$ asymptotics feasible, one has to choose a very special parametrization of matrix  $U\in K$.
Inspired by Helgason's derivation in the nonsupersymmetric case \cite{Helgason} and previous experience for supersymmetric models \cite{Eslam_thesis}, we find it appropriate to factor $U$ as
\color{black}
\be
\label{U_as_three_matrices}
  U = U_{\text{BB}} U_{\text{FF}} U_g ,
\ee
where
\begin{gather}
  \quad U_{\text{BB}}
  =
  e^{i  \alpha_{b1} w_{b1}} e^{i  \beta_{b1} w_{b2}} e^{i \alpha_{b2} w_{b1}} e^{i \beta_{b2} w_{b2}} ,
\nonumber \\
\label{Ug_for_three_parametric}
  U_{\text{FF}} = e^{i \alpha_F w_F} ,
\\ \nonumber
	U_g = e^{w_{g1}} e^{w_{g2}} = (1 + w_{g1}) (1 + w_{g2}) ,
\end{gather}
the generators $w_{b1}, w_{b2}, w_F, w_{g1}, w_{g2}$ are a sum of a pair of opposite root vectors defined in Appendix~\ref{app:basis_and_root_system}, and $\alpha_{b1},  \beta_{b1}, \alpha_{b2},  \beta_{b2}, \alpha_F$ are real numbers,
which belong to the domains
$\alpha_{b2}, \beta_{b2} \in [0, \pi]$ and $\alpha_{b1}, \beta_{b1}, \phi_F \in [0, 2 \pi]$ [the domains follow from the position of the singular points of the Jacobian \eqref{Jacobian_for_U}].
The most delicate part is to parametrise the BB sector,  $U_\text{BB}$. It appears that the proper way (allowing to obtain tractable integrals for the large-$\theta_i$ asymptotics) is to act with the two generators, formed from the so-called \textit{simple roots} in an alternating way (see Ref.~\onlinecite{Helgason} and Appendix~\ref{app:basis_and_root_system}).

In this parametrization the Haar measure
for the group $K$ corresponds to the following Jacobian
\be
\label{Jacobian_for_U}
	J_U = \sin \alpha_{b2} \sin^2 \beta_{b2} .
\ee
It is this factorization of the Jacobian $J_U$ into the product of simple trigonometric functions that along with the similar integrand structure following from Eq.\ \eqref{equating_minors} allows one to calculate the asymptotic expression of the wavefunctions $\phi_\nu(\theta)$ in the limit $\theta_{B1} \gg \theta_{B2} \gg 1$ in an explicit form, see Section \ref{SS:measure}.
The parametrization \eqref{U_as_three_matrices} and \eqref{Ug_for_three_parametric} will be used below to obtain the principal, three-parametric family of radial eigenfunctions.

The general theory of noncompact symmetric spaces \cite{Helgason} suggests that momenta $p_i$ in Eq.\ \eqref{plane-wave} should be shifted by the Weyl vector \eqref{Weyl_vector}, $p_i = q_i + (i/2) \rho$, in order to obtain normalisable wavefunctions at real $q_i$.
In our case we perform such a shift for the BB sector, while for the FF sector we use the parametrization that is convenient to obtain the wavefunctions that behave correctly at $\thF = \pi$ (see discussion in Sec.\  \ref{SSS:asymptotical_expressions}). Namely, we reparametrise the Iwasawa momenta $p_\text{B1}, p_\text{B2}, p_\text{F}$ in the following way:
\be
\label{momenta_reparametrization}
	p_\text{B1} = q_1, \qquad
   p_\text{B2} = q_2 - i/2,  \qquad
   p_\text{F} = - i l.
\ee
Then proper eigenfunctions are then parametrised by $q_1 \geq q_2 \geq 0$ and $l=0,1,\dots$ (see Secs.\ \ref{SSS:asymptotical_expressions} and \ref{SSS:three-param}).

For noncompact symmetric spaces, the described procedure yields the complete basis of radial eigenfunctions \cite{Helgason}. In contrast, in the case of supersymmetric spaces, additional subfamilies of eigenfunctions do emerge. We will discuss them below.

\subsection{Families of radial eigenfunctions}
\label{SS:eigenfunctions}

An explicit expression for three-parametric functions $\phi_{q_1 q_2 l}(\theta)$ is unknown as the integral \eqref{wavefunction_K_integral} over the group $K$ cannot be calculated for arbitrary $\theta_i$. Nevertheless it is possible to demonstrate that it vanishes at a special line $\thtwo = \thF = 0$ due to the fact that not more than six out of eight Grassmann variables are present in every monomial of the integrand. The same is true for the line $\thone = \thF = 0$ due to the Weyl group symmetry. However, it does not belong to the chosen Weyl chamber $\thone \geq \thtwo \geq 0$ and thus should not be considered. Mentioned nullification means that three-parametric radial functions along with the unit function 1 do not constitute a  complete basis. So, for example, $\delta(Q,\Lambda)$ cannot be expanded in $\phi_{q_1 q_2 l}(\theta)$ and 1 at least at the mentioned line, which is the first arising issue.

The second issue is that putting $q_2 = l = 0$ and leaving only $q_1$ also nullifies the integral \eqref{wavefunction_K_integral} for the same reason, which may indicate that wavefunctions corresponding to these momenta are lost. In order to recover the lost eigenfunctions we will consider the $q_2 = l = 0$ family in a modified parametrization and will omit integration over some Grassmann variables. As a result, we will also resolve the first mentioned issue on incompletness of the basis of three-parametric eigenfunctions.

The modification of the parametrization should make every term in the integrand of \eqref{wavefunction_K_integral} lack the same subset of Grassmann variables, so that we can omit integration over them in the process of izotropization. We achieve that by using the parametrization \eqref{U_as_three_matrices} and \eqref{Ug_for_three_parametric}, but with $U_g$ replaced by
\be
\label{Ug_for_one_parametric}
	\tilde{U}_g = (1 + \left. w_{g2} \right|_{\gamma,\chi \rightarrow 0}) (1 + w_{g1}) (1 + \left. w_{g2} \right|_{\rho,\sigma \rightarrow 0}) .
\ee
That modifies the Jacobian \eqref{Jacobian_for_U}: $J_U \rightarrow J_U (1 + 4 \eta \chi - 4 \zeta \gamma)$ and essentially makes the whole integrand in Eq.\ \eqref{wavefunction_K_integral} independent of four Grassmann variables $\alpha, \beta, \rho, \sigma$. Omitting integration over these four variables in accordance with the general logic of izotropization, we arrive at an additional family of one-parametric eigenfunctions $\phi_{q_1}(\theta)$.

To sum up, for class D with two replicas, there exist three families of radial eigenfunctions:
\begin{itemize}
\item three-parametric functions $\phi_{q_1 q_2 l}(\theta)$, which are obtained by averaging over the full group $K$
(vanish at the ``bosonic line'' $\thtwo=\thF=0$);

\item one-parametric functions $\phi_{q_1}(\theta)$, arising when a plane wave in Iwasawa coordinates does not depend on some Grassmann variables and integration over them is not performed (izotropization);
remarkably, this family is closely related to the eigenfunctions of the transfer-matrix Hamiltonian for the sigma-model of class D with one replica, as we show below
(vanish at the origin $\thone=\thtwo=\thF=0$);

\item unit function 1, corresponding to the trivial plane wave 1 in Iwasawa coordinates, which should not be integrated over Grassmann variables at all.
\end{itemize}

This situation is to be contrasted with the case of class D with one replica, when the only nontrivial one-parametric family $\phi_q(\theta_\text{B})$ can be obtained in a standard way by averaging over the full group $K$ \cite{BagretsKamenev}.

We suppose that there are no other eigenfunctions of the Laplacian. We check this statement in Appendix~\ref{sec:basis_check}.

The eigenvalues of the obtained wavefunctions are given by \eqref{eig-Iw} and equal:
\be
\label{eigenvalues}
	\epsilon_{q_1} = q_1^2, \quad \epsilon_{q_1 q_2 l} = \frac{1}{4} + q_1^2 + q_2^2 + \frac{1}{2} l (l + 1) .
\ee

As mentioned above, the functions $\phi_{q_1 q_2 l}(\theta)$ and $\phi_{q_1}(\theta)$ cannot be obtained in a closed form. Resorting to computer algebra system, we are able to calculate their values explicitly only in some particular cases, where two of three Cartan angles $\theta_i$ are set to zero (``bosonic'' and ``fermionic'' lines) and in the large-$\theta$ asymptotic regime. Below we present behavior of radial functions on these lines and discuss their asymptotic behavior at large and small $\theta$.

Expressions for these particular cases will be sufficient to calculate physical observables (conductance and its variance): large-$\theta_i$ asymptotics allows to determine integration measure $\mu_\nu$ in Eq.~\eqref{heat_kernel} (see Sec.~\ref{SS:measure}), while values at the ``fermionic line'' will be used to determine the overall numerical coefficient, check the heat kernel construction and conveniently obtain small-$\theta_i$ expansion. The values at the ``bosonic line'' simplify determination of the measure for one-parametric wavefunctions and illuminate connection between $n=2$ and $n=1$ sigma-models.

\subsubsection{``Bosonic line'' $(\thtwo = \thF = 0)$}
\label{SSS:bosonic_line_values}

Three-parametric radial wavefunctions vanish at the ``bosonic line'': $\phi_{q_1 q_2 l}(\thone,0,0) = 0$.

As the ``bosonic line'' contains only one BB angle $\thone$ (like in the $n=1$ case) and one-parametric functions depend only on one momentum $q_1$ (like in the $n=1$ case), one may expect that at the ``bosonic line'' the sigma-model with two replicas reduces to the sigma-model with one replica.
Such a reduction indeed takes place, and it can be proved by studying the action of the Laplacian on the wave function. To this end,
we substitute the expansion of the wavefunction in the vicinity of the ``bosonic line'' ($\thtwo \ll1$, $\thF \ll 1$),
\be
\phi = f(\thone) + u(\thone) \thtwo^2 + v(\thone) \thF^2 + \dots ,
\ee
into Eq.~\eqref{eig} with the radial Laplacian given by Eq.~\eqref{radial_laplacian} and obtain
\be
\label{Laplacian_on_expansion_near_bosonic_line}
	\Delta \phi = \Delta^{(1)} f(\thone) + \frac{\thtwo^2 - \thF^2}{\thtwo^2 + \thF^2} \left[ 2v(\thone) - 4 u(\thone) \right] .
\ee
Here $\Delta^{(1)}$ is the one-replica radial Laplacian \cite{BagretsKamenev, Eslam_thesis},
\be
  \Delta^{(1)}
  =
  \frac{1}{J^{(1)}} \frac{\partial}{\partial \thone}
  J^{(1)} \frac{\partial}{\partial \thone} ,
\ee
with the one-replica Jacobian
$J^{(1)}(\thone) = \coth({\thone}/{2})$,
which can be obtained from Eq.\ \eqref{Jacobian} by sending $\thtwo$ and $\thF$ to zero and omitting singular $\thone$-independent factors.
Equation \eqref{Laplacian_on_expansion_near_bosonic_line} being substituted into Eq.\ \eqref{eig} indicates that the $\thtwo = \thF = 0$ limit of the eigenfunction $\phi$ is well-defined only if $v = 2u$. Then the last term drops and equation for $f(\thone)$ acquires a form of the Laplace operator in class D with only one replica.
Hence we can readily identify the eigenfunctions in this limit \cite{BagretsKamenev, Eslam_thesis}:
\be
\label{1P_at_bosonic_line}
  \phi_{q_1} (\thone, 0, 0)
  =
  i q_1 \left[P_{i q_1} (\lambda_1)
  -  P_{-i q_1} (\lambda_1) \right],
\ee
where $\lambda_1 = \cosh \thone$ and $P_\nu (z)$ is the Legendre function. The corresponding eigenvalues are given by Eq.\ \eqref{eigenvalues}.
The wavefunctions \eqref{1P_at_bosonic_line} are orthogonal when integrated over the ``bosonic line'' with the Jacobian $J^{(1)}(\thone)$.

In the asymptotical region $\thone \gg 1$ one-parametric eigenfunctions $\phi_{q_1}$ behave as
\be
\label{asymptotics_1P}
	\phi_{q_1} \sim c_{q_1}    e^{i q_1 \thone},
\ee
with the coefficient $c_{q_1}$ (Harish-Chandra $c$-function)
\be
\label{Harish_Chandra_1P}
	c_{q_1} = i q_1 C_{q_1}, \qquad C_{q_1} =   \frac{1}{\sqrt{\pi}} \frac{\Gamma(1/2 + i q_1)}{\Gamma(i q_1)} .
\ee
This function is used to obtain the integration measure $\mu_{q_1}$ in the heat kernel \eqref{heat_kernel_specific} [see Eq.~\eqref{mu_1P_through_c}].

\color{black}

\subsubsection{``Fermionic line'' $(\thone = \thtwo = 0)$}

Three-parametric radial eigenfunctions on the ``fermionic line''  $\thone = \thtwo = 0$ can be obtained by taking the integral \eqref{wavefunction_K_integral} with the help of a computer algebra system. That requires processing $\sim 3500$ terms, each of them integrated via the formula
\be
	\int_0^{2\pi} d \phi \left( \cos \theta + i \sin \theta \cos \phi \right)^\nu = 2\pi P_{\nu} (\cos \theta).
\ee
Using identities for the Legendre function allows us to bring the obtained expression to a compact form:
\begin{multline}
\label{fermionic_line_3P}
\phi_{q_1 q_2 l}(0,0,\thF)
= 16 (l^2 + 4 q_1^2)(l^2 + 4 q_2^2)
	  P_{l}(\lambda_F) \sin^4 \frac{\thF}{2}
\\
	  {} +  32 (1 + l)  \epsilon_{q_1 q_2 l}
	 \left[ P_{l}(\lambda_F) - P_{1 + l} (\lambda_F) \right] \sin^2 \frac{\thF}{2} ,
\end{multline}
where $\lambda_F = \cos \thF$.

For one-parametric functions at the ``fermionic line'' we get:
\be
\label{fermionic_line_1P}
	 \phi_{q_1}  (0,0,\theta_F) = - 4 q_1^2 \sin^2 \frac{\thF}{2} .
\ee

\subsubsection{Asymptotic behavior at $\thone \gg \thtwo \gg 1$}
\label{SSS:asymptotical_expressions}

In the limit $\thone \gg \thtwo \gg 1$,
three-parametric wavefunctions behave as
\be
\label{asymptotics_with_Legendre}
	\qquad \phi_{q_1 q_2 l} \sim \mathcal{W} \, \tilde{c}_{q_1 q_2 l}  \cdot P_{l} (\lambda_F)   e^{i q_1 \thone + (i q_2 + 1/2) \thtwo},
\ee
with
\be
\label{Harish_Chandra_3P_tilde}
	\tilde{c}_{q_1 q_2 l} =  \frac{(1 + l - 2 i q_1) (l + 2 i q_1) (1 + l - 2 i q_2) (l + 2 i q_2)} {\pi^4 C_{q_1} C_{q_2} C_{q_1 + q_2} C_{q_1 -  q_2} } ,
\ee
where the coefficients $C_q$ are defined in \eqref{Harish_Chandra_1P}.
In Eq.\ \eqref{asymptotics_with_Legendre}, the operation $\mathcal{W}$ denotes symmetrization with respect to the BB Weyl symmetry group \cite{Helgason}, namely for arbitrary function $F_{q_1, q_2}$ of variables $q_1$, $q_2$:
\be
\label{Weyl_group_symmetrization}
	\qquad \mathcal{W} \, F_{q_1, q_2}
  =
  \sum_{\sigma_1,\sigma_2=\pm1}
  \bigl( F_{\sigma_1 q_1, \sigma_2 q_2} + F_{\sigma_1 q_2, \sigma_2 q_1} \bigr),
\ee
where the sum is taken over all possible sign choices [four for each term in \eqref{Weyl_group_symmetrization}].

Expression \eqref{fermionic_line_3P} implies that for the wavefunction to be well-defined at $\thF = \pi$, Legendre function should reduce to Legendre polynomial at integer $l$. Taking into account that  $P_l(z) = P_{-1-l}(z)$, we conclude that
the allowed discrete momenta are
$l = 0, 1, 2, \dots$

\color{black}

\subsubsection{Behavior at small $\theta_i$}

Small-$\theta_i$ expansion of the wavefunctions can be regularily obtained from the integral \eqref{wavefunction_K_integral}, but we find it more convenient to expand expression  \eqref{fermionic_line_3P} for the wavefunctions on the fermionic line and then use relations \eqref{small_theta_relations_first} for the Taylor series coefficients, which follow from the symmetry properties of the action \eqref{sigma_model_action} and Schr\"odinger equation \eqref{Schroedinger_equation}, see Appendix~\ref{app:small_th} for the derivation.

\subsection{Eigenfunctions normalization and Plancherel measure}
\label{SS:measure}

Having identified the families of radial eigenfunctions, we can rewrite the general expression \eqref{heat_kernel} for the heat kernel in an explicit form:
\begin{multline}
\label{heat_kernel_specific}
	\psi(Q,x) = 1 + \int_{-\infty}^\infty dq_1 \, \mu_{q_1} \phi_{q_1}(\theta) e^{-2\epsilon_{q_1} {x}/{\xi}}
\\{}
	 + \int_{-\infty}^\infty dq_1 dq_2 \sum_{l = 0}^{\infty} \mu_{q_1q_2 l} \phi_{q_1 q_2 l}(\theta) e^{-2\epsilon_{q_1 q_2 l} {x}/{\xi}},
\end{multline}
where $\mu_{q_1}$ and $\mu_{q_1q_2 l}$ are integration measures that will be determined below. A few comments are in order here.
First, the unit eigenfunction enters with the coefficient 1 in order to respect the boundary condition \eqref{supersymmetric_delta_function} at $x\to0$ since all other eigenfunctions vanish at the origin ($\theta_i = 0$).
Second, as the three-parametric eigenfunctions are symmetric with respect to the Weyl symmetry group (interchange and sign flip of $q_1$, $q_2$, see Sec.~\ref{SS:eigenfunctions}), each eigenfunction in Eq.\ \eqref{heat_kernel_specific} is actually taken several times. However, as the integration measure $\mu_{q_1 q_2 l}$ also obeys the same property, we prefer to keep integration over all $q_1$ and $q_2$, adjusting the overall numerical factor in $\mu_{q_1 q_2 l}$.
Third, the formulas for $\mu_{q_1}$ and $\mu_{q_1 q_2 l}$ that we present below are written for the particular choice of the overall normalization coefficient of the one- and three-parametric families specified in  Eqs.~\eqref{1P_at_bosonic_line}--\eqref{asymptotics_1P} and \eqref{fermionic_line_3P}--\eqref{fermionic_line_1P}.
 Fourth, strictly speaking we do not have a proof that the eigenfunctions 1, $\phi_{q_1}(\theta)$ and $\phi_{q_1, q_2 l}(\theta)$ do form a basis and no other radial eigenfunctions exist. However a strong evidence of that is provided by the numerical check that Eq.\ \eqref{heat_kernel_specific} indeed reproduces the supersymmetric delta-function \eqref{supersymmetric_delta_function} at $x\to0$, see Appendix~\ref{sec:basis_check}.
Another strong argument in favor of the correctness of the heat kernel \eqref{heat_kernel_specific}
is that the average conductance,
its variance and the third cumulant calculated from it in the small-$L$ limit coincide with the perturbative results obtained in Appendix \ref{app:perturbative_Laplacian}. This fact is rather nontrivial since it requires cancellation of $1/L^2$ and $1/L$ terms in the series expansion for $\var g$ and five leading terms ($1/L^3$ through $L$) for $\corr{\corr{g^3}}$.

For noncompact symmetric spaces without Grassmann variables, the integration (Plancherel) measure $\mu$ is determined by asymptotic behavior of wavefunctions \cite{Helgason}. It is given by $\mu_q^{\text{noncomp}} = \text{const}/|c_{q}|^2 = \text{const}/(c_{q} c_{-q})$, where the \textit{Harish-Chandra $c$-function} $c_q$ is a coefficient in the large-$\theta$ asymptotics of the wavefunctions obtained with the help of Iwasawa parametrization [see Eq.~\eqref{wavefunction_K_integral}]. For supersymmetric spaces, the strict mathematical proof is lacking, however it is generally believed that the analogous formula, originally proposed by Zirnbauer \cite{Zirnbauer1991, Zirnbauer1992} still works.

As in our convention the wavefunctions  \eqref{wavefunction_K_integral} are defined up to an arbitrary overall numerical factor, their normalization should be consistent with the integration measure. The latter will be determined in the process of numerical check of the basis completeness on the ``fermionic line''
in Appendix~\ref{sec:basis_check} with the help of Eqs.\ \eqref{fermionic_line_3P} and \eqref{fermionic_line_1P} .

\subsubsection{One-parametric eigenfunctions, $\phi_{q_1}(\theta)$}

For one-parametric eigenfunctions the generalization of the noncompact-case expression for the measure is rather straightforward:
\be
\label{mu_1P_through_c}
	\mu_{q_1}  = \frac{1}{2 \pi} \frac{1}{c_{q_1} c_{-q_1}} = \frac{\coth \pi q_1}{2 q_1} ,
\ee
which is valid provided that $\phi_{q_1}(\theta)$ is normalized such that its behavior at the ``bosonic line'' is given by Eq.\ \eqref{1P_at_bosonic_line} with the Harish-Chandra $c$-function \eqref{Harish_Chandra_1P}.

Equation \eqref{mu_1P_through_c} can be completely inherited from the $n=1$ sigma model \cite{BagretsKamenev, Eslam_thesis}, since on bosonic line the eigenbasis completely turns to the eigenbasis of the $n=1$ sigma-model as discussed in Sec.~\ref{SSS:bosonic_line_values}.

\subsubsection{Three-parametric eigenfunctions, $\phi_{q_1q_2l}(\theta)$}
\label{SSS:three-param}

For three-parametric wavefunctions the generalization of the nonsupersymmetric formula for the measure is a bit more intricate. The suggested procedure \cite{Zirnbauer1991} is the following.

First we need to consider completely noncompact theory by taking analytically continued asymptotics at big negative imaginary $\thF$. Substituting $\thF = -i \vartheta_F$ and using asymptotical behaviour of the Legendre function at large argument one gets that at $\thone \gg \thtwo \gg \vartheta_F \gg 1$ the wavefunction, accompanied by the $\sqrt{J}$ factor behaves as:
\be
\label{asymptotics_analytically_continued}
	\sqrt{J} \phi_{q_1 q_2 l} \sim c_{q_1 q_2 l} \mathcal{W}  \, e^{i q_1 \thone + i q_2 \thtwo + (1/2 + l)  \vartheta_F},
\ee
which corresponds to a normalisable wavefunction at real $q_1$, $q_2$ and $l = -1/2 + l_F$ with  imaginary $l_F$.
Harish-Chandra $c$-function is given by
\be
\label{Harish_Chandra_3P}
	c_{q_1 q_2 l} = \frac{\tilde{c}_{q_1 q_2 l}}{\pi c_{-i(l+1/2)}},
\ee
where $\tilde{c}_{q_1 q_2 l}$ was defined in \eqref{Harish_Chandra_3P_tilde}.

Applying usual formula for the Plancherel measure to this noncompact theory gives
\be
\label{measure_analytically_coninued}
	\mu_{q_1 q_2 l}^{\text{noncomp}} = \frac{\text{const}}{|c_{q_1 q_2 l}|^2} =
	\frac{\text{const}}{c_{q_1 q_2 l} c_{-q_1, -q_2, -1 - l} } .
\ee
In the original theory with real $\thF$ proper values of $q_1$ and $q_2$ (corresponding to normalisable wavefunctions) are also real, which justifies \eqref{momenta_reparametrization}. Proper $l$ were derived from \eqref{asymptotics_with_Legendre} (see the discussion there) and are given below \eqref{momenta_reparametrization}. It appears that \eqref{measure_analytically_coninued} has poles in these values and the proposed formula for the measure is:
\be
\label{measure_as_residue}
	\mu_{q_1 q_2 l} = \res_{l} \frac{1}{c_{q_1 q_2 l} c_{-q_1, -q_2, - 1 - l} }, \quad l = 0, 1,\dots ,
\ee
if the overall numerical factor in the wavefunction is chosen according to \eqref{fermionic_line_3P}.
We get the numerical coefficient $1/\pi^4$ in \eqref{measure_as_residue} and justify this formula in Appendix~\ref{sec:basis_check}.

Substituting \eqref{Harish_Chandra_3P} to \eqref{measure_as_residue} we get the integration measure in the form:
\be
	\mu_{q_1 q_2 l} = \frac{(1 + 2l) T_{q_1} T_{q_2} T_{q_1 + q_2} T_{q_1 - q_2}}{2 \prod_{q\in\{ q_1, q_2\}} [l^2 + 4 q^2]  [(1+l)^2 + 4 q^2]} ,
\ee
where $T_q = q \tanh \pi q$ .

\section{Analytical expressions for transport characteristics}
\label{sec:results}

Now we are in position to compute quasiparticle transport properties of the superconducting wire in class D.
This is done by substituting the partition function \eqref{Z_eq_psi}
expressed via the heat kernel \eqref{heat_kernel_specific} into Eqs.~\eqref{G_through_Z}--\eqref{var_det_through_Z}. With the help of relations \eqref{small_theta_relations_first}, the emerging $\theta$ derivatives can be expressed in terms of derivatives only over $\theta_\text{F}$. This allows to consider the wavefunctions $\phi_{q_1q_2l}(\theta)$ and $\phi_{q_1}(\theta)$ only at the ``fermionic line'', where they are given explicitly by Eqs.\ \eqref{fermionic_line_3P} and \eqref{fermionic_line_1P}.

\black

\subsection{Conductance moments}
\label{sec:results_moments}
This procedure yields the following expressions for the average conductance, its second and third moments:
\begin{multline}
\label{result_general}
 \langle g^k \rangle = \int_{-\infty}^\infty dq \, \mu_{q} P_{q}^{(k)}  e^{-2\epsilon_{q} {L}/{\xi}}  \\
  + \int_{-\infty}^\infty dq_1 dq_2 \sum_{l = 0}^{\infty} \mu_{q_1q_2 l}
R_{q_1q_2l}^{(k)}
e^{-2\epsilon_{q_1 q_2 l} {L}/{\xi}},
\end{multline}
where the eigenvalues $\epsilon_{q}$ and $\epsilon_{q_1 q_2 l}$ are listed in \eqref{eigenvalues}, while the measures $\mu_{q_1}$ and $\mu_{q_1q_2l}$ are given by Eqs.\ \eqref{mu_1P_through_c} and \eqref{measure_as_residue}, with the Harish-Chandra $c$-functions specified in Eqs.\ \eqref{Harish_Chandra_1P} and \eqref{Harish_Chandra_3P}, respectively.  The polynomials $P_{q_1}^k$ defining the contribution of one-parametric eigenfunctions have the form:
\begin{subequations}
\label{P-polynomials}
\begin{align}
	& P_{q}^{(1)} = 4 q^2,
\label{P-polynomials-1}
\\
	& P_{q}^{(2)} = \frac{8}{3} q^2 (1 + q^2),
\label{P-polynomials-2}
\\
	& P_{q}^{(3)} = \frac{8}{15} q^2 (1 + q^2) (4 + q^2).
\label{P-polynomials-3}
\end{align}
\end{subequations}
The contribution of three-parametric eigenfunctions is described by the polynomials $R_{q_1q_2l}^{(k)}$:
\begin{subequations}
\label{R-polynomials}
\begin{align}
	& R_{q_1q_2l}^{(1)} = 0,
\label{R-polynomials-1}
\\
	& R_{q_1q_2l}^{(2)} = \frac{64}{3} \left[4 \epsilon_{q_1 q_2 l} (1 + l)^2 + M \right],
\label{R-polynomials-2}
\\
	&
R_{q_1q_2l}^{(3)} = \frac{32}{5} \left[
	4 \epsilon_{q_1 q_2 l} (1+l)^2  B_{3}
+ M B_{4}
	\right],
\label{R-polynomials-3}
\end{align}
\end{subequations}
where
$M=(l^2 + 4 q_1^2)(l^2 + 4 q_2^2)$ and
$B_m = 5 + l (4 + m l) + 4 q_1^2 + 4 q_2^2$.

The obtained expression for the average conductance $\corr{g}$ coincides with that calculated from the one-replica sigma model \cite{BagretsKamenev, Eslam_thesis} (with account for different normalization of $g$ \cite{com-factors}).
Note, however, that we obtain it from the analysis of a more complicated two-replica sigma model. Therefore this
anticipated
coincidence can be considered as a consistency check of our treatment of the $n=2$ case. Mathematically, the fact that three-parametric functions $\phi_{q_1q_2l}(\theta)$ do not contribute to $\corr{g}$, but contribute to $\corr{g^2}$ and $\langle g^3 \rangle$
is a consequence of the fact that their Taylor expansion at small $\theta_i$ does not contain quadratic terms, starting with quartic terms [see Eq.~\eqref{fermionic_line_3P}].

The asymptotic behavior of the conductance moments \emph{in the long-wire limit}, $L \gg \xi$, is determined by the first term in Eq.~\eqref{result_general}, as the three-parametric spectrum is gapped, while the one-parametric spectrum is not [see Eq.~\eqref{eigenvalues}].
Evaluating the integral over $q_1$ with the steepest descent method, we find
\begin{subequations}
\label{g-gg-long}
\begin{align}
\label{g1-long}
	& \frac{\gavg}{g_L} =
1 + \frac{\pi^2 \xi}{12 L} - \frac{\pi^4 \xi^2}{240 L^2} + \dots ,
\\
	& \frac{\langle g^2 \rangle}{g_L} =
\frac{2}{3} + \frac{(3 + \pi^2) \xi}{18 L} + \frac{\pi^2 (15 - \pi^2) \xi^2}{360 L^2} + \dots ,
\label{g2-long}
\\
	& \frac{\langle g^3 \rangle}{g_L} = \frac{8}{15} + \frac{(15 + 4\pi^2) \xi}{90 L} + \dots ,
\label{g3-long}
\end{align}
\end{subequations}
where $g_L=\sqrt{2\xi/\pi L}$.
The leading asymptotics in Eqs.\ \eqref{g-gg-long} conform with the result of the DMPK approach \cite{GruzbergReadVishveshwara} and can be obtained from the Dorokhov distribution \eqref{one_channel_Dorokhov}.

\emph{In the short-wire limit}, $L \ll \xi$, both subfamilies of the radial eigenfunctions contribute to $\corr{g^2}$ and $\corr{g^3}$.
However it is known to be easier to extract analytical expressions for the short-wire asymptotics not from the general formula \eqref{result_general}, but from the direct perturbative solution of the Schr\"odinger equation for the heat kernel \cite{MMZ}, which is done in Appendix~\ref{app:perturbative_Laplacian}. Alternatively, one can evaluate the fluctuational determinant near the quasiclassical trajectory \cite{Eslam_thesis}.
We find numerical result to be consistent with both analytical methods, that provides a strong evidence of the correctness of the basis construction. The resulting expansions for $\corr{g}$, $\var g = \langle g^2 \rangle - \gavg^2$ and $\corr{\corr{ g^3}} = \langle g^3 \rangle - 3 \langle g^2 \rangle \langle g \rangle + 2 \langle g \rangle^3$ read:
\begin{subequations}
\label{g-com-short}
\begin{align}
\label{g1-short}
	& \gavg = \frac{\xi}{L} + \frac{1}{3} - \frac{1}{15} \frac{L}{\xi} + \frac{2}{63} \frac{L^2}{\xi^2} + \dots,
\\
\label{g2-short}
	& \var g = \frac{2}{15} - \frac{8}{315} \frac{L}{\xi} + \frac{136}{4725}\frac{L^2}{\xi^2} + \dots ,
\\
\label{g3-short}
	& \corr{\corr{g^3}} = \frac{8}{1485} \frac{L^2}{\xi^2} + \dots
\end{align}
\end{subequations}
\black
In the process of $\var g$ calculation,
two leading
terms proportional to $1/L^2$ and $1/L$ completely cancel, as expected for universal conductance fluctuations \cite{UCF_Lee_Stone}.
Surprisingly, the leading term for the third cumulant is proportional to $L^2$ rather than $L$, as would follow from the scaling $\langle \langle g^k \rangle \rangle \propto L^{k-2}$ suggested in Ref.\ \cite{AKL}. Such cancellation of the leading contribution to $\corr{\corr{g^3}}$ in the weak-localization regime is known to be a peculiar feature of the one-dimensional geometry \cite{vanRossum1997}.

At arbitrary wire length $L$, the average conductance and its variance should be calculated numerically. The results are presented in
Figs.~\ref{fig:conductance}, \ref{fig:variance} and \ref{fig:cumGGG}, which illustrate the crossover from the Drude regime at small $L$ to the critical regime at large $L$.
Quite unexpectedly, $\var g$ and $\corr{\corr{g^3}}$ approach their asymptotic limits \eqref{g-com-short} much slower than $\gavg$ itself.

\subsection{ Fano factor}
\label{sec:shot-noise}

The pseudo-Fano factor \eqref{pseudo-Fano} given by Eq.\ \eqref{F_through_Z} can also be cast in the form of Eq.\ \eqref{result_general}, with the polynomials $P$ and $R$ replaced by
\black
\be
\label{Q_P_ShotNoise}
	P_q^{\text{(F)}} = \frac{4}{3} q^2 -\frac{8}{3} q^4, \qquad   R_{q_1q_2l}^{\text{(F)}} = \frac{1}{2} R_{q_1q_2l}^{(2)} .
\ee
The resulting dependence of the pseudo-Fano factor $\tilde F$ on the wire length is shown in Fig.\ \ref{fig:QFano}. Its large- and small-$L$ asymptotics are given by
\be
\label{Fano_asymptotics}
  \tilde{F}
  =
  \begin{cases}
    \displaystyle
    \frac{1}{3} - \frac{1}{6} \frac{\xi}{L} - \frac{\pi^2}{36} \frac{\xi^2}{L^2} + \dots,
    & L \gg \xi ,
\\[9pt]
    \displaystyle
    \frac{1}{3} - \frac{4}{45} \frac{L}{\xi} + \frac{76}{945} \frac{L^2}{\xi^2} + \dots,
    & L \ll \xi .
  \end{cases}
\ee
\black
As mentioned in Introduction, coinciding asymptotic values of $1/3$ are explained by the fact that both limits are described by the bimodal Dorokhov function \eqref{one_channel_Dorokhov}, giving the total density of many transmission eigenvalues at small $L$ (Drude regime, self-averaging Fano factor) and the distribution function of one most transparent channel at large $L$ (critical regime, strong fluctuations).

\subsection{Variance of $\det r$}

The variance \eqref{var_det_through_Z} of the determinant of the reflection amplitudes matrix $\var \det r = \langle \text{det}^2 r \rangle$ is expressed via the heat kernel at the ``south pole'' and does not involve $\theta_i$-derivatives.
Hence it contains the unit contribution from the unity eigenfunction [the first term in Eq.\ \eqref{heat_kernel_specific}],
while the contribution of one- and three-parametric eigenfunctions is given by \eqref{result_general} with the polynomials
\be
	P_q^{(\text{det})} = -4 q^2, \qquad   R_{q_1q_2l}^{(\text{det})} = \frac{3 (-1)^l}{4} R_{q_1q_2l}^{(2)} .
\ee
The dependence of $\var\det r$ on the wire length is shown in Fig.\ \ref{fig:ChiChi}. Its large- and small-$L$ asymptotics have the form
\be
\label{var_det_asymptotics}
  \var\det r
  =
  \begin{cases}
    \displaystyle
    1 -  \sqrt{\frac{2 \xi}{\pi L}} + \dots,
    & L \gg \xi .
\\[9pt]
    \displaystyle
    \sim
    \exp\left( -\frac{\pi^2}{4} \frac{\xi}{L} \right),
    & L \ll \xi .
  \end{cases}
\ee

The property $\lim_{L\to\infty}\var\det r=1$ means that even in the critical regime most of the samples demonstrate insulating behaviour, being deep either in the topological or trivial phases.
As $\var \det r$ is determined by the heat kernel at the south pole \eqref{var_det_through_Z}, the fact that it vanishes in the limit $L \rightarrow 0$ is yet another check of correctness of our heat kernel construction.

\color{black}

\section{Conclusion}
\label{sec:conclusion}

In the present paper, we performed an extensive study of quasiparticle transport in disordered multichannel ($N \gg 1$) quantum wires of symmetry class D, which can be implemented in superconductors with broken time-reversal and spin-rotation symmetries, where quasiparticles determine thermal rather than electrical conductance. This symmetry class allows for two distinct topological phases, depending on the parameters of the Hamiltonian. At large lengths both phases are subject to Anderson localization, while the critical regime realised at the boundary between the two phases demonstrates a peculiar ``delocalization'' behavior, in which average transport properties are determined by rare configurations, described by the Dorokhov distribution for the most transparent channel.

The average conductance $\corr{g}$ in quantum wires of class D was calculated in Refs.\ \cite{Zirnbauer1991, BagretsKamenev, Eslam_thesis} in the framework of the nonlinear supersymmetric sigma model with one replica $(n=1)$. This approach allows to describe the full dependence of $\corr{g}$ on the wire length $L$ (see Fig.\ \ref{fig:conductance}), tracing the crossover from the common behaviour $\gavg = \xi / L$ in the Drude regime ($L \ll \xi$) to the super-Ohmic behaviour $\gavg \propto \sqrt{\xi / L}$ in the critical regime ($L \gg \xi$), where $\xi = 2 N l$ is the correlation length of the wire.

In our work, we make a next step towards full statistical description of quantum transport in class D and generalized previous studies by  calculating higher-order moments of the conductance: its variance and the third cumulant. Extracting these quantities requires the use of a more complicated nonlinear supersymmetric sigma model with two replicas ($n = 2$), which has never been analyzed before for the symmetry class D, to the best of our knowledge. The supersymmetric sigma model with two replicas is defined on the symmetric supermanifold of rank three (i.e.\ with three Cartan angles), making it possible to access conductance moments up to the third order. Interestingly, our results for both $\var g$ (Fig.\ \ref{fig:variance}) and $\corr{\corr{g^3}}$ (Fig.\ \ref{fig:cumGGG}) demonstrate a broad crossover region and approach their long-wire limit only at $L \gtrsim 20\xi$. At the same time, the average conductance (Fig.\ \ref{fig:conductance}) is well described by its asymptotic expression already at $L \gtrsim \xi$.

The $n=2$ sigma model analyzed in the present work is also suitable for describing the full distribution of transmission probabilities and hence allows to extract the full counting statistics (FCS) of the wire. The distribution of transmission probabilities can be expressed in terms of the heat kernel in the vicinity of the ``supersymmetric line'' $\thone = \thtwo = - i \thF$. The peculiarity of the symmetry class D is that the FCS generating function \eqref{generating_function} cannot be deduced from the $n=1$ sigma model, whose compact sector is essentially empty and the corresponding Cartan angle is lacking. Therefore the theory with $n=2$ is the minimal model for extracting the FCS. A very complicated structure of the integral representation of the eigenfunctions \eqref{wavefunction_K_integral} based on the Iwasawa decomposition prevents us from direct analytical calculation of the distribution of transmission probabilities. However, individual moments of this distribution can be written in a concise form. This includes the average conductance and Fano factor  \eqref{Q_P_ShotNoise}. The latter approaches its quasiclassical value $1/3$ both in the short- and long-wire limits, see Eq.\ \eqref{Fano_asymptotics} and Fig.\ \ref{fig:QFano}.

Finally, we calculate the variance of the determinant of the matrix of reflection amplitudes (see Fig.~\ref{fig:ChiChi}). This determinant is related to the topological index $\chi = \sign\det r$ of the wire and defines the transition between the two topologically distinct localized phases. Throughout the paper we considered the critical state of the wire for which the determinant is zero on average. At the same time, the average square of the determinant has a non-trivial dependence on the wire length. It indicates that at $L \gg \xi$ most of the samples undergo Anderson localization, while the probability to find a conducting wire decreases as $\sqrt{\xi / L}$.

Average quasiparticle conductance and its moments can be accessed via heat flow
measurements as in Ref.~\cite{SET_heat_transport}. An alternative experiment can address electrical
(rather than thermal) shot noise power in response to the applied temperature
gradient. This kind of a measurement in a non-superconducting sample was
discussed in Refs.~\cite{Lumbroso_experiment, Sivre_experiment}. 
Electrical shot noise in a superconducting system is
very different from the thermal shot noise considered in the present paper
(see Fig.~\ref{fig:QFano}). Particular relations between electrical noise and scattering
properties of a superconducting sample will be the subject of a separate
publication.

Mesoscopic fluctuations of the transport properties can be studied on a single sample by varying some external parameters such as magnetic field or gate voltage. This variation should be performed according to a special protocol to keep the system at the critical state between the two topologically distinct phases. Such a sweep will perform an effective averaging over disorder realizations and allow to gain the necessary statistics.

From a technical perspective, our calculation is based on the construction of the full set of eigenfunctions of the radial Laplace-Beltrami operator on the sigma-model supermanifold of class D with two replicas. This task is accomplished by using the Iwasawa decomposition of the corresponding supergroup $G$ and subsequent averaging of the radial plane waves in Iwasawa coordinates with respect to the rotations by the subgroup $K$.
This approach was first proposed in Refs.\ \cite{Zirnbauer1992,MMZ}
and applied there to the minimal (one replica) models of the standard Wigner-Dyson classes.

We have observed that for the supersymmetric sigma model of class D with two replicas there are exist two distinct subfamilies of eigenfunctions aside from the special zero mode
(identically unity on the whole manifold).
One generic eigenfunction family is parametrized by three components of momentum,
in accordance with the presence of three Cartan angles. The peculiarity of class D with two replicas is that
all these eigenfunctions identically vanish on the special ``bosonic line'' $\thtwo = \thF = 0$. A smaller one-parameter subfamily of eigenfunctions remains finite on this line and is intimately related to the eigenfucnctions of the model with one replica. The latter model has only one Cartan angle corresponding to $\thone$  and ``lives'' exactly on the ``bosonic line''. Interestingly, the spectrum of the one-parameter subfamily is gapless contrary to the three-parametric set of eigenfunctions. Hence most properties of the wire in the limit $L \gg \xi$ are dominated by the one-parameteric subfamily.

It is instructive to compare our analysis of radial eigenfunctions of the sigma model for class D with two replicas with that for the orthogonal (AI) and symplectic (AII) symmetry classes in the one-replica case \cite{Zirnbauer1992,MMZ}. The target spaces of all these sigma models have rank 3, with three Cartan angles in each case. Moreover, the one-parametric subfamily we identified for $n=2$ class D is partially reminiscent of the ``subsidiary series'' eigenfunctions for $n=1$ classes AI and AII. The principle difference however is that in our case additional eigenfunctions cannot be obtained by taking certain limits of the main three-parametric eigenfunction family and strictly speaking cannot be derived by a naive application of the Iwasawa trick. Instead, averaging over the $K$ group should be understood as isotropization, when integration over some Grassmann variables should be discarded if they do not explicitly appear in the integrand. Such a complication is a consequence of the supersymmetry and does not arise in the theory of conventional symmetric spaces.

From the structure of our results we conclude that
such a hierarchical organization of eigenfunctions
is generic and applies to supersymmetric sigma models of all classes with an arbitrary number $n > 1$ of replicas. Namely, the full set of eigenfunctions in each of these models includes as special subsets eigenfunctions of the model with fewer replicas 
(properly extended to a manifold with a larger dimensionality).
The special unit eigenfunction that exists in sigma models of all classes and is constant (independent of all Cartan angles) can be also viewed as such a special subset corresponding to the model with zero replicas.

One interesting possible extension of our results include quantum wires with topologically protected channels. Physically, this corresponds to edge transport in 2D  topological insulators and superconductors. Wires of symmetry classes A, C, and D can host any integer number of protected channels that corresponds to the $\mathds{Z}$ topology. In classes AII and DIII the topological index is $\mathds{Z}_2$ that corresponds to a single protected channel in the case when the total number of channels is odd. The presence of topologically protected channels leads to the appearance of a Wess-Zumino-Witten (WZW) term in the sigma-model action and modifies the spectrum of corresponding eigenfunctions. Quasiclassical description of wires with protected channels was developed in Ref.\ \cite{KSO2016}. The full set of eigenfunctions for the unitary class A with the WZW term was constructed in Ref.\ \cite{KO2017} with the help of Sutherland transformation. Iwasawa decomposition of the supermanifold can be also used to construct eiegenfunctions of the models with the WZW term. This will be the subject of a separate publication \cite{classD_topology} both for wires of class D and other symmetry classes with protected channels.

\black

\acknowledgments

We are grateful to A.~Gorsky, I.~Gruzberg, B.~Halperin, A.~Kamenev and O.~Motrunich for stimulating discussions and to C.~W.~J. Beenakker for his illuminating comment on the nature of the shot noise in Majorana systems.
This work was partially supported by the Russian Science Foundation under Grant No.\ 20-12-00361.

\appendix
\section{Notations, basis, etc.}

\subsection{Basis and the root system}
\label{app:basis_and_root_system}

We use the basis in which bosonic and fermionic sectors are selected according to the grading matrix
\be
	k = \diag \{ 1, 1, -1, -1, -1, -1, 1, 1 \} ,
\ee
which acts as $1$ on bosons and $-1$ on fermions.

The origin (``north pole'') $\Lambda$ and charge conjugation matrix $C$ are chosen in the form
\begin{gather}
\label{Lambda-def}
	\Lambda =
	\begin{pmatrix}
		0 & 0 & 0 & 0 & 0 & 0 & 0 & 1 \\
		0 & 0 & 0 & 0 & 0 & 0 & 1 & 0 \\
		0 & 0 & 0 & 0 & 0 & 1 & 0 & 0 \\
		0 & 0 & 0 & 0 & 1 & 0 & 0 & 0 \\
		0 & 0 & 0 & 1 & 0 & 0 & 0 & 0 \\
		0 & 0 & 1 & 0 & 0 & 0 & 0 & 0 \\
		0 & 1 & 0 & 0 & 0 & 0 & 0 & 0 \\
		1 & 0 & 0 & 0 & 0 & 0 & 0 & 0
	\end{pmatrix},
\\
\label{C-def}
	C =
	\begin{pmatrix}
		0 & 0 & 0 & 0 & 0 & 0 & 0 & 1 \\
		0 & 0 & 0 & 0 & 0 & 0 & 1 & 0 \\
		0 & 0 & 0 & 0 & 1 & 0 & 0 & 0 \\
		0 & 0 & 0 & 0 & 0 & -1 & 0 & 0 \\
		0 & 0 & 1 & 0 & 0 & 0 & 0 & 0 \\
		0 & 0 & 0 & -1 & 0 & 0 & 0 & 0 \\
		0 & -1 & 0 & 0 & 0 & 0 & 0 & 0 \\
		-1 & 0 & 0 & 0 & 0 & 0 & 0 & 0
	\end{pmatrix} .
\end{gather}
Then the Cartan algebra can be parametrised as follows:
\be
\label{check_theta}
	\check{\theta} = \diag \{ \thone, \thtwo, i \thF, i \thF, -i \thF, -i \thF, -\thtwo, -\thone \} .
\ee

The metrics $\mathfrak{g}_{\alpha \beta}$ on the sigma-model supermanifold is defined via the length element \cite{Efetov-book, MMZ, Eslam_thesis}
\be
\label{dl_appendix}
	dl^2 = -\frac{1}{2} \str dQ^2 = \mathfrak{g}_{\alpha \beta} dX^\alpha dX^\beta .
\ee
The radial Laplacian \eqref{radial_laplacian} is determined by the $\theta$-de\-pen\-dent part of $\mathfrak{g}$. Plugging Efetov's parametrization \eqref{Cartan_parametrization} into Eq.\ \eqref{dl_appendix}, we obtain the radial part of the length element (which appears to be $U$-independent):
\be
	dl_{\text{rad}}^2 = -\frac{1}{2} \str (\Lambda e^{\check{\theta}} d \check \theta)^2 = \frac{1}{2} \str d \check{\theta}^2 .
\ee
Taking $\check\theta$ from Eq.\ \eqref{check_theta}, we arrive at Eq.~\eqref{g_metrics}, which defines the radial part of the metrics $g_{ij}$.
Note, however, that the definition of the Laplace-Beltrami operator \eqref{Hamiltonian_is_Laplacian} contains the upper-index metrics $g^{ij} = (g_{ij})^{-1}$. This is the reason why in Eq.\ \eqref{radial_laplacian} the coefficient in the $\thF$-derivative term is $1/2$ rather than $2$ as in Eq.\ \eqref{g_metrics}. The same matrix $g^{ij}$ defines the dot product for roots (dual Cartan space), thus entering Eqs.~\eqref{Iwasawa_laplacian} and \eqref{eig-Iw}.

\begin{figure}
	\includegraphics[width=0.85\linewidth]{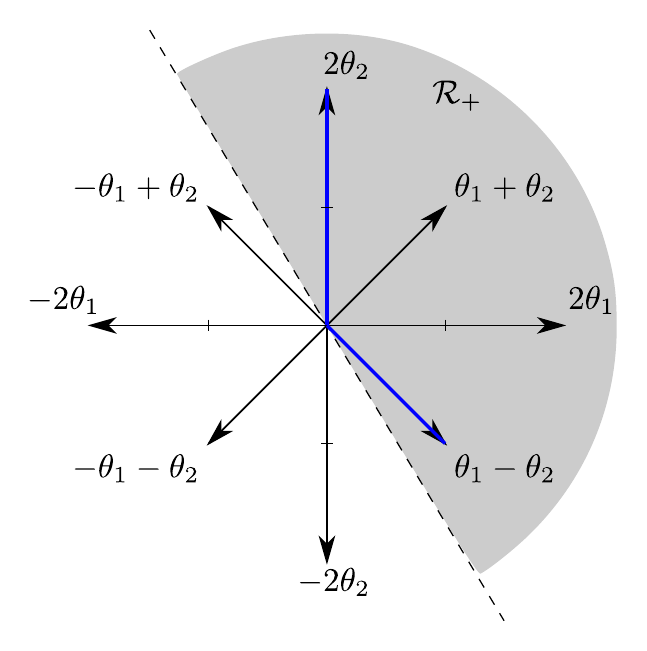}
	\protect\caption{Boson-boson part of the root system for the supermanifold of $n=2$ sigma-model. Chosen half-plane $\mathcal{R}_{+}$, containing positive roots is show in grey. Simple roots are highlighted with bold blue colour.
}
\label{fig:roots}
\end{figure}

A crucial advantage of the chosen basis is that it allows to choose positive root vectors so that they are upper triangular matrices. We summarise thus selected positive roots in Table \ref{T:roots} and depict its BB part in Fig.~\ref{fig:roots}.

\begin{table}
\caption{Root system: positive roots ($\alpha$), their multiplicities ($m_\alpha$) and corresponding root vectors ($Z_{\alpha(,i)}$).
The matrix $\Xi_{ij}$ has 1 at the position $(i,j)$ and 0 elsewhere.
}
\label{T:roots}
\begin{ruledtabular}
\begin{tabular}{ccccc}
\multicolumn{2}{c}{Bosonic ($m_\alpha=1$)} &
\multicolumn{3}{c}{Fermionic ($m_\alpha=-2$)} \\
\hline
$\alpha$ & $Z_{\alpha}$ &
$\alpha$ & $Z_{\alpha,1}$ & $Z_{\alpha,1}$ \\
\hline
$2 \thone$ & $\Xi_{18}$ &
$\thone + i \thF$ & $\Xi_{15}-\Xi_{38}$ & $\Xi_{16}+\Xi_{48}$ \\
$2 \thtwo$ & $\Xi_{27}$ &
$\thone - i \thF$ & $\Xi_{13}-\Xi_{48}$ & $\Xi_{14}+\Xi_{68}$ \\
$2 \thF$   & $\Xi_{36} + \Xi_{45}$ &
$\thtwo + i \thF$ & $\Xi_{25}-\Xi_{37}$ & $\Xi_{26}+\Xi_{47}$ \\
$\thone + \thtwo$ & $\Xi_{17} - \Xi_{28}$ &
$\thtwo - i \thF$ & $\Xi_{23} - \Xi_{57}$ & $\Xi_{24} + \Xi_{67} $ \\
$\thone - \thtwo$ & $\Xi_{12} - \Xi_{78}$
&&& \\
\end{tabular}
\end{ruledtabular}
\end{table}

In the BB sector, a special role is played \cite{Helgason} by the so-called \emph{simple roots}\  $2 \thtwo$ and $\thone - \thtwo$ (depicted by blue bold arrows in Fig.~\ref{fig:roots}). These are the roots that lie most closely to the boundary of the chosen half-plane $\mathcal{R}_{+}$ in Cartan space containing positive roots. In other words, all positive roots can be expressed as a linear combination of simple roots with positive coefficients. Acting with exponentials of these roots in an alternating way [see Eq.~\eqref{Ug_for_three_parametric}] allows to construct a parametrization, in which the integral \eqref{wavefunction_K_integral} can be analytically taken in the limit of large $\theta_i$, giving an explicit expression for the Harish-Chandra function \eqref{Harish_Chandra_3P_tilde}.

\subsection{Generators of the $K$ group}

We parametrise the $K$ group (see Sec.\ \ref{sec:sigma-model definition}) by the generators $w_{\alpha}$ that are formed as a sum of a root vector $Z_{\alpha}$ and its counterpart $Z_{-\alpha}$, corresponding to the opposite root $-\alpha$ . In other words, we make a $\Lambda$-commuting matrix from each positive root:
\be
	w_{\alpha} = Z_{\alpha} + \Lambda Z_{\alpha} \Lambda .
\ee
In Sec.~\ref{sec:eigenfunctions_constructions} we use short-cut notations $w_{b1} = w_{2\theta_{B2}}$, $w_{b2} = -i w_{\theta_{B1} - \theta_{B2}}$, and $w_F = w_{2\theta_F}$ for the generators of the $U$ group in the BB and FF sectors, and the following Grassmann generators:
\begin{gather}
	w_{g1} =
	\begin{pmatrix}
		0 & 0 & \eta & \zeta & \zeta & \eta & 0 & 0 \\
		0 & 0 & \alpha & \beta & \beta & \alpha & 0 & 0 \\
		\zeta & \beta & 0 & 0 & 0 & 0 & -\beta & -\zeta \\
		-\eta & -\alpha & 0 & 0 & 0 & 0 & \alpha & \eta \\
		\eta & \alpha & 0 & 0 & 0 & 0 & -\alpha & -\eta \\
		-\zeta & -\beta & 0 & 0 & 0 & 0 & \beta & \zeta \\
		0 & 0 & \alpha & \beta & \beta & \alpha & 0 & 0 \\
		0 & 0 & \eta & \zeta & \zeta & \eta & 0 & 0
	\end{pmatrix},
\\
	w_{g2} =
	\begin{pmatrix}
		0 & 0 & \gamma & \chi & -\chi & -\gamma & 0 & 0 \\
		0 & 0 & \rho & \sigma & -\sigma & -\rho & 0 & 0 \\
		\chi & \sigma & 0 & 0 & 0 & 0 & -\sigma & -\chi \\
		-\gamma & -\rho & 0 & 0 & 0 & 0 & -\rho & -\gamma \\
		-\gamma & -\rho & 0 & 0 & 0 & 0 & -\rho & -\gamma \\
		\chi & \sigma & 0 & 0 & 0 & 0 & \sigma & \chi \\
		0 & 0 & -\rho & -\sigma & \sigma & \rho & 0 & 0 \\
		0 & 0 & -\gamma & -\chi & \chi & \gamma & 0 & 0
	\end{pmatrix} .
\end{gather}

\section{Transport properties and heat kernel}
\label{app:transport_via_SM}

Transport properties of a quasi-one-dimensional wire can be characterised by a set of transparency coefficients that are eigenvalues of the matrix $t^\dagger t$, where $t$ is the matrix of transmission amplitudes. Full statistics of transmission coefficients \cite{LL1993, LLYa1995} can be conveniently encoded in the generating function
\be
\label{generating_function}
	\mathcal{F}(z) = \sum_{k=1}^{\infty} z^{k - 1} \tr (t^\dagger t)^k = \tr \frac{t^\dagger t}{1 - z t^\dagger t}.
\ee
In particular, the
dimensionless
conductance and zero-frequency shot noise power \cite{Landauer, Buttiker, Landauer_IBM} described by the Fano factor $F$ can be extracted via \begin{subequations}
\label{g-Fano}
\begin{gather}
	\label{g_via_F}
		g = \tr t^\dagger t = \mathcal{F} (0), \\
	\label{SN_via_F}
		g F = \tr t^\dagger t  - \tr (t^\dagger t)^2 =  \mathcal{F} (0) - \mathcal{F}^{\prime}_{z} (0) .
\end{gather}
\end{subequations}

In order to compute $\mathcal{F}(z)$, Nazarov \cite{Nazarov_PRL} introduced a special matrix Green function, where the standard retarded and advanced functions are mixed by an auxiliary counting field. Translated to the sigma-model language \cite{MMZ, KSO2016, Eslam_thesis}, Nazarov's counting field appears in the twisted boundary conditions for the sigma model.  The disorder-averaged generating function is then expressed as
\be
\label{F_SUSY}
	\langle \mathcal{F}(z_\text{F}) \rangle = - \left. \frac{\partial Z[\theta_i]}{\partial z_\text{F}} \right|_{\text{SUSY line}},
\ee
where $Z[\theta_i]$ is the partition function \eqref{partition_function},
\be
\label{generating_function_via_SUSY_line}
	z_\text{F} = \sin^2 (\thF/2) ,
\ee
and the derivative in Eq.\ \eqref{F_SUSY} should be taken at the ``supersymmetric line'', where the fermionic and bosonic angles are equal.

The symmetry class D considered in the present paper has several features to be taken into account in the general scheme outlined above. First, when studying quasiparticle properties at zero energy, Nambu-Gor'kov space plays the role of the retarded-advanced space of Nazarov's matrix. Second, in the one-replica ($n=1$) case, the FF sector of the sigma-model supermanifold is degenerate and lacks the corresponding Cartan angle $\thF$. This makes it impossible to construct the full generating function $\corr{\mathcal{F}(z)}$. Instead, only the value at the origin, $\corr{\mathcal{F}(0)}$, is accessible, giving the average conductance  \cite{BagretsKamenev, Eslam_thesis} via Eq.\ \eqref{G1_through_Z}.

As was explained in Introduction, the $n=2$ sigma model possesses two bosonic ($\thone, \thtwo$) and one fermionic ($\thF$) Cartan angles. This is sufficient to apply Eq.\ \eqref{F_SUSY} and obtain the complete FCS generating function. The supersymmetric line in this case corresponds to $\thone = \thtwo = -i \thF$.

Computation of the full generating function from the heat kernel requires the knowledge of the eigenfunctions of the Laplace-Beltrami operator in the vicinity of the supersymmetric line. The integral representation \eqref{wavefunction_K_integral} based on the Iwasawa decomposition turns out to be too complicated for this task. Direct calculation of the eigenfunctions in this limit is not feasible for an arbitrary value of $\thF$. However, the moments of the distribution can be directly accessed by expanding the eigenfunctions in small vaues of all three Cartan angles as explained in Appendix \ref{app:small_th}. For instance, substituting Eq.\ \eqref{F_SUSY} into Eq.\ \eqref{SN_via_F} we derive the Fano factor in the form of Eq.\ \eqref{F_through_Z}.

\black
It is worth noting that due to the presence of three Cartan angles, the average conductance can be extracted from the partition function \eqref{partition_function} in three different ways: (i) from the fermionic sector with the help of Eqs.\ \eqref{F_SUSY} and \eqref{g_via_F}:
\be
	\left\langle g \right\rangle = -2 \left.\frac{\partial^{2}Z(\theta_i)}{ \partial \thF^2}\right|_{0},
\ee
(ii) from the bosonic sector through Eq.\  \eqref{G1_through_Z}, and (iii) from its counterpart with $\thone \rightarrow \thtwo$.
The existence of the three copies of the systems that are jointly averaged over disorder opens a way to go beyond the linear statistics and to calculate the second and the third moments of the conductance given by Eqs.\ \eqref{G2_through_Z} and \eqref{G3_through_Z}, respectively.

\section{Small $\theta$ expansion of radial wavefunctions}
\label{app:small_th}

According to Eqs.\ \eqref{G_through_Z} and \eqref{F_through_Z}, conductance moments and Fano factor are expressed in terms of derivatives of the heat kernel, and hence of the eigenfunctions, at the origin. In this Appendix we demonstrate that these derivatives can be expressed via derivatives taken along the special ``fermionic line'' $\thone = \thtwo = 0$, where the eigenfunctions are known explicitly, see Eqs.~\eqref{fermionic_line_3P} and \eqref{fermionic_line_1P}.
The relations that we derive apply both to three-parametric eigenfunctions $\phi_{q_1 q_2 l}$ and to one-parametric eigenfunctions $\phi_{q_1}$ [the corresponding eigenvalues $\epsilon_{q_1 q_2 l}$ and $\epsilon_{q_1}$ are presented in Eq.~\eqref{eigenvalues}] and
allow to calculate physical quantities given by Eqs.\ \eqref{G_through_Z} and \eqref{F_through_Z}.

The symmetry of the action \eqref{sigma_model_action} and, hence, the Laplacian \eqref{radial_laplacian} with respect to the rotations by the $K$ group
[see Sec.\ \ref{sec:sigma-model definition}
for definition]
implies that the small-$\theta$ expansion of a radial eigenfunction $\phi$ should be expressed via $K$-invariant polynomials
\be
\label{H_polynomials}
	H_n = \frac{1}{2} \str \check{\theta}^n =
 \thone^{2n} + \thtwo^{2n} - 2(-\thF^2)^n
\ee
in the form of a series (constant term drops in all eigenfunctions except the unity)
\be
\label{small_theta_general}
	\phi(\theta_i) = a_1 H_1 + a_2 H_2 + a_{1,1} H_1^2 +  \dots
\ee

Substituting Eq.\ \eqref{small_theta_general} into Eq.~\eqref{eig} with the Laplace operator \eqref{radial_laplacian}
determined by the Jacobian \eqref{Jacobian} and expanding in $\theta_i$ to the sixth order, one obtains a number of relations between the coefficients:
\begin{gather}
	(1 + 3 \epsilon) a_1 + 24 (a_2 + a_{1,1}) = 0,
\nonumber
\\
	\left( \frac{2}{3} + \epsilon \right) a_2 - \frac{a_1}{45} + 16 a_{1,2} + 18 a_3 = 0,
\nonumber
\\
	\left( \frac{8}{3} + 4 \epsilon \right) a_{1,1} - \frac{a_1}{15} + 96 a_{1,1,1} + 32 a_{1,2} + 24 a_3 = 0 .
\nonumber
\end{gather}

The general form of an eigenfunction \eqref{small_theta_general} together with the above relations allows to express different derivatives of $\phi$ via $\thF$-derivatives and eigenvalue $\epsilon$ only:%
\begin{subequations}
\label{small_theta_relations_first}
\begin{gather}
\label{small_theta_relation_first}
 	\left.\frac{\partial^{2} \phi}{ \partial \thone^2}\right|_{0} =
		\frac{1}{2} \left.\frac{\partial^{2} \phi}{ \partial \thF^2}\right|_{0}, \\
	\left.\frac{\partial^{4} \phi}{ \partial \thone^4}\right|_{0} =
		- \frac{1 + 3 \epsilon}{4} \left.\frac{\partial^{2} \phi}{ \partial \thF^2}\right|_{0}, \\
\label{small_theta_relation_second}
 	\left.\frac{\partial^{4} \phi}{ \partial \thone^2  \partial \thtwo^2} \right|_{0} =
		- \frac{3\epsilon + 1}{36} \left.\frac{\partial^{2} \phi}{ \partial \thF^2}\right|_{0} + \frac{1}{18} \left.\frac{\partial^{4} \phi}{ \partial \thF^4}\right|_{0} , \\
	\left.\frac{\partial^{6} \phi}{ \partial \thone^2  \partial \thtwo^2 \partial \thF^2} \right|_{0} =
		\frac{15 \epsilon (1 + \epsilon) + 16}{1800} \left.\frac{\partial^{2} \phi}{ \partial \thF^2}\right|_{0}
\hspace{20mm}
		 \nonumber \\
\hspace{25mm}{}  - \frac{2 + 3 \epsilon}{90} \left.\frac{\partial^{4} \phi}{ \partial \thF^4}\right|_{0}
		+ \frac{1}{450} \left.\frac{\partial^{6} \phi}{ \partial \thF^6}\right|_{0}
  .
\end{gather}
\end{subequations}
\black

\section{Completeness of the eigenfunction set}
\label{sec:basis_check}

Integral representation \eqref{heat_kernel_specific} of the heat kernel is based on the expansion of unity in the eigenfunctions of Laplace-Beltrami operator. While we do not have a direct proof of the completeness of our basis, in this Appendix we will provide a numerical evidence that the expansion \eqref{heat_kernel_specific} with the weights defined by Eqs.\ \eqref{mu_1P_through_c} and \eqref{measure_as_residue} indeed reproduces the full heat kernel. First, the one-parameter family of eigenfunctions and their weights are fixed by the expansion of unity on the ``bosonic line'' $\thtwo = \thF = 0$. Hence the first two terms of Eq.\ \eqref{heat_kernel_specific} are beyond any doubts. In order to demonstrate correctness of the third term involving three-parametric family of eigenfunctions, we will consider the heat kernel on the ``fermionic line'', where $\thone = \thtwo = 0$. Let us note that small $\theta$ expansion of the eigenfunctions constructed in Appendix \ref{app:small_th} is fully determined by their behavior on the fermionic line. Hence completeness of the heat kernel on this line is sufficient to ensure that our results for conductance moments and Fano factor are correct.

We consider the heat kernel \eqref{heat_kernel_specific} on the ``fermionic line'', where the eigenfunctions are known explicitly [Eqs.\ \eqref{fermionic_line_3P} and \eqref{fermionic_line_1P}].
Using recurrence relations for Legendre polynomials, we represent the three-parameter eigenfunction $\phi_{q_1,q_2,l}$ as a linear combination of up to five terms $P_s(\cos\thF)$ with orders $l - 2 \leq s \leq l + 2$. Similarly, the one-parameter eigenfunction is a linear combination of $P_0$ and $P_1$. Using this representation in Eq.\ \eqref{heat_kernel_specific}, we collect the terms with the same Legendre polynomial:
\begin{multline}
 \label{fermionic_line_heat_kernel}
 \psi(0,0,\thF; x) = 1 + \sum_{s=0}^{\infty} \biggl[ \int dq_1 dq_2 A^{(s)}_{q_1,q_2}(x)
\\{} + \int dq \, B^{(s)}_{q}(x) \biggr] P_s(\cos \thF).
\end{multline}
Explicit forms of the coefficients $A^{(s)}$ and $B^{(s)}$ are rather lengthy but unimportant. As we just explained above, $B^{(s)}_q$ is nonzero only for $s = 0$ or $1$, while $A^{(s)}_{q_1,q_2}$ is present for all values of $s$.

In the limit $x = 0$, the heat kernel represents a supersymmetric delta function \eqref{supersymmetric_delta_function} and should vanish for all nonzero values of $\thF$.
This means that all terms of the above sum with $s \geq 1$ must be zero, while the $s = 0$ term compensates the unity. We checked and confirmed this statement numerically for the terms up to $s = 6$. It is worth noting that the two contributions with double and single momentum integrals in Eq.\ \eqref{fermionic_line_heat_kernel} diverge in the limit $x \to 0$ as $1/x$, while their sum remains finite. This happens for $s = 0$ and $1$ when both terms exist. In this case, we have to compute integrals for several small values of $x$ in order to cancel the divergence and single out the finite value. For $s \geq 2$ only the double integral remains. This integral converges on its own and yields $0$. It can be calculated directly at $x = 0$.

\section{Direct perturbative computation of the heat kernel}
\label{app:perturbative_Laplacian}

As mentioned in Sec.~\ref{sec:results_moments}, it is hard to extract analytical formulas for transport characteristics in the short-wire limit $L \ll \xi$ from the Iwasawa trick \eqref{result_general}. However, this regime can be easily accessed via direct perturbative calculation of the heat kernel $\psi(\theta_i, L)$ from the Schr{\"o}dinger equation \eqref{Schroedinger_equation} \cite{MMZ}.
For this aim, we substitute expression \eqref{J_through_roots} for the Jacobian via the root system into the radial Laplacian \eqref{radial_laplacian} and rewrite it in the form \be
\label{Laplacian_through_coth}
	\Delta_\text{rad} = \partial \cdot \partial + \sum_{\alpha \in R^+} m_\alpha \coth \alpha \,\partial_{(\alpha)},
\ee
where $\partial_i = \partial / \partial \theta_i$, $\partial_{(\alpha)} = (\partial_i \alpha) g^{ij} \partial_j$ denotes the derivative in the direction $\alpha$, and the dot product is defined with respect to the metrics $g^{ij}$ [see Eq.\ \eqref{g_metrics}]. Positive roots $\alpha \in R^+$ are listed in Table~\ref{T:roots}: There are five bosonic roots with multiplicity $m_\alpha = 1$ (four BB roots are depicted in Fig.~\ref{fig:roots}) and four Grassmann roots (with two root vectors for each) that are to be counted with multiplicity $m_\alpha = -2$ \cite{MMZ}.

The idea behind a perturbative construction of the heat kernel $\psi(\theta_i, L)$ is that for  $L\ll\xi$ relevant $\theta$ are small, and the Laplacian can be well approximated by its Euclidean version acting in the tangent plane:
\be
\label{flat_Laplacian}
	\Delta_\text{rad}^\text{E} = \partial \cdot \partial + \sum_{\alpha \in R^+} m_\alpha \alpha^{-1} \,\partial_{(\alpha)} ,
\ee
where both terms scale as $\theta^{-2}$. If we now replace $\Delta_\text{rad}$ by $\Delta_\text{rad}^\text{E}$, the Schr\"odinger equation \eqref{Schroedinger_equation} can be easily solved, providing the Euclidean approximation to the heat kernel:
\be
\label{perturbative_zero_order}
	\psi^\text{E}(\theta,L) = \exp( - \xi H_1/8L ) ,
\ee
where $H_1 = \thone^2 + \thtwo^2 + 2 \thF^2$ is the first in the family of invariant polynomials \eqref{H_polynomials}.
The absence of a usual prefactor $\propto L^{-d/2}$ in the heat kernel \eqref{perturbative_zero_order} is a consequence of the supersymmetry of the theory, which makes it effectively zero-dimensional, $d=0$.

To improve the approximation \eqref{perturbative_zero_order}, one has to take into account the curvature of the sigma-model manifold, i.e. the difference between $\Delta_\text{rad}$ and $\Delta_\text{rad}^\text{E}$. For symmetry reasons, the heat kernel is expressed only in terms of the invariant polynomials \eqref{H_polynomials}. Writing it as a series
\be
\label{perturbation_series}
	\psi = e^{-\xi H_1/8L} \left[1 + b_1 H_1 + b_2 H_2 + b_{1,1} H_1^2 + \dots \right]
\ee
with $L$-dependent coefficients $b_i$ and expanding the Schr\"odinger equation in $\Delta_\text{rad}-\Delta_\text{rad}^\text{E}$,
one can extract short-wire asymptotics order by order:
\begin{subequations}
\label{pertubation_result}
\begin{gather}
b_1 = -\frac{1}{24} + \frac{1}{120} \frac{L}{\xi} - \frac{1}{252} \frac{L^2}{\xi^2} + \dots, \\
b_2 = \frac{1}{720} - \frac{1}{1260} \frac{L}{\xi} + \dots , \\
b_{1,1} = \frac{11}{5760} - \frac{11}{20160} \frac{L}{\xi} + \dots
\end{gather}
\end{subequations}
Substitution the perturbative heat kernel \eqref{perturbation_series} to Eqs.\ \eqref{G_through_Z} and \eqref{F_through_Z} provides analytical expressions for the short-wire asymptotics of physical quantities given in Sec.~\ref{sec:results}.

\bibliography{class_D_variance}

\begin{thebibliography}{62}
\bibitem [{com()}]{com-factors}%
  \BibitemOpen
  \href@noop {} {}\bibinfo {note} {The coefficient $\xi/16$ in the sigma-model
  action \eqref{sigma_model_action} is the same in Ref.\ \cite{BagretsKamenev}
  (where it is denoted by $\tilde\xi$), Ref.\ \cite{Eslam_thesis}, and the
  present publication. The difference in the results should be probably
  attributed to a different choice of the conductance quantum. Our
  dimensionless conductance $g$ is 4 times larger than that in
  Ref.~\cite{BagretsKamenev} and 2 times larger than that in
  Ref.~\cite{Eslam_thesis}}\BibitemShut {NoStop}%
\end{thebibliography}

\end{document}